\documentclass{aa}  

\usepackage{pdflscape}

\usepackage{graphicx}
\usepackage{txfonts}
\begin{document}

   \title{Extreme Emission-Line Galaxies in the MUSE Hubble Ultra Deep Field Survey}

   \author{I. del Moral-Castro\inst{\ref{inst1},\ref{inst2},\ref{inst3}}\thanks{E-mail: ignaciodelmoralcastro.astro@gmail.com}
          \and
          J. M. V\'ilchez\inst{\ref{inst4}}
          \and
          J. Iglesias-Páramo\inst{\ref{inst4},\ref{inst5}}
          \and
          A. Arroyo-Polonio\inst{\ref{inst4}}
          }

    \institute{Instituto de Astrofísica, Facultad de Física, Pontificia Universidad Católica de Chile, Campus San Joaquín, Av. Vicuña Mackenna 4860, Macul, Santiago, Chile, 7820436 \label{inst1}
                \and
                Kapteyn Astronomical Institute, University of Groningen, PO Box 800, 9700 AV Groningen, The Netherlands \label{inst2}
                \and
                Facultad de F\'isica, Universidad de La Laguna, Avda. Astrof\'isico Fco. S\'anchez s/n, 38200, La Laguna, Tenerife, Spain\label{inst3}
                \and
                Instituto de Astrof\'isica de Andaluc\'ia – CSIC, Glorieta de la Astronom\'ia s/n, 18008 Granada, Spain\label{inst4}
                \and
                Centro Astronómico Hispano en Andalucía, Observatorio de Calar Alto, Sierra de los Filabres, 04550 Gérgal, Spain\label{inst5}
                }

   \date{Received   ; accepted }

 
  \abstract
   {}
   {We apply a methodology to build a sample of extreme emission line galaxies (EELGs) using integral field spectroscopy data. In this work we follow the spectroscopic criteria corresponding for EELG selection and use the MUSE Hubble Ultra-Deep field survey, which includes the deepest spectroscopic survey ever performed. 
   }
   {Objects in the primary (extended) sample were detected requiring a rest-frame equivalent width EWo $\geqslant$ 300 \AA{} (200\AA{} $\leq$ EWo $\leq$ 300\AA{}) in any of the emission lines of [OII]$\lambda\lambda$3726,29,  [OIII]$\lambda\lambda$5007,4959, or H$\alpha$. A detailed closer inspection of the spectra of the candidates selected has been performed on a one by one basis, in order to confirm their classification. For this sample, the line fluxes, physical properties and chemical abundances of the EELGs have been derived as well as their spatially resolved structure and kinematics.}
   {Four (five) of the galaxies in the primary (extended) sample, $\sim$57$\%$ ($\sim$83$\%$),  were spatially resolved. Three (none) of them present  a clear pattern compatible with rotation. We have shown how our entire EELGs sample share the same loci defined by high-redshift galaxies (z $\approx$ 6-8) for the mass-metallicity relation, illustrating their role as local analogs.}
   {}


   \maketitle
%

\section{Introduction}\label{Sec:Intro}

Galaxies hosting strong events of star formation can provide keys for our understanding of early galaxy evolution, since such extreme star-forming objects are expected to be more frequent towards the first epochs of the Universe. In this sense, these highly star-forming galaxies provide useful windows to the high-z Universe, allowing us to study the photon budget required for its re-ionization (e.g. \citealt{Stark_2016,Dayal_2018}). The nature, properties and prevalence of these galaxies at the Epoch of Reionization (EoR) remain to be fully understood.

All these galaxies have been named as a class as extreme emission line galaxies (EELGs) since their (optical) spectra show very strong emission lines with extreme equivalent width (EW). EELGs host a considerable number of massive young star forming systems producing a substantial amount of photoionising radiation. This radiation is absorbed by the surrounding metal-poor gas (e.g., \citealt{Ravindranath_2020}). Since the gas is expected to have a perforated irregular distribution, hard ionising radiation could escape (e.g. \citealt{Bergvall_2006, Izotov_2016, Fletcher_2019, Perez_montero_2020}). It turns these
systems into perfect laboratories for the sutdy of LyC escape. Therefore, the discovery and characterisation of local EELG analogs give us a way to gain more insight about first moments of galaxy evolution when leakage of LyC photons from EELGs provided a major contribution to the EoR photon budget (e.g. \citealt{Erb_2016,2017ApJ...844..171Y,Naidu_2022,Matthee_2022}).

EELGs present spectra with very high EW of [OIII] and H$\beta$ emission lines, and it has been found that these galaxies use to be also strong Ly$\alpha$ emitters \citep{Tang_2021}. On the other hand, most typical LyC leakers presenting a very high EW(Ly$\alpha$) are frequently found among Green Pea galaxies, an outstanding family among the prototypes of EELGs \citep{Izotov_2016,Izotov_2018,Perez_montero_2021}.

As of today, observational studies have discovered thousands of EELGs but most searches have produced a mixed bag of denominations, typically inspired by their compact optical appearance and/or colour in plates; examples of these are Green Peas, Blue Berries, ELDots/H$\alpha$Dots, or SDSS extreme HII galaxies and their cohorts, to cite a few (e.g. \citealt{Terlevich_1991,Cardamone_2009,Amorin_2010,AmorinVilchez2012_rotating,Bekki_2015,Yang_2017,Salzer_2020}), all  illustrating how EELG selection can depend on the redshift range explored and the exact criteria and technique applied. The search for EELGs is somewhat complementing the historical searches for blue galaxies of very compact morphology that were identified and analyzed in pioneering works since the middle of the last century (e.g. \citealt{Haro_1956,Zwicky_1966,Markarian_1967}).  However, it is important to emphasize here that a large majority of EELG searches performed so far were based on broad-band photometry data, thus avoiding direct spectroscopic or narrow-band based selection based on emission lines which still remains scarce (see e.g. \citealt{Indahl_2021, Lumbreras-Calle_2022,Jorge_2022_EELG_JPAS}). This situation may have somehow supported the persistence of systematics and/or obvious bias in current samples of EELG, given the typical very faint continua and the strong emission lines of these galaxies. A clear example are the differences in the number and occurrence of these objects. \citet{Paalvast_2018} explored galaxies with extreme [OIII]/[OII] ratios combining data from several MUSE Guaranteed Time Observing (GTO) programmes (0.28 $<$ z $<$ 0.85) founding 15 of these systems in their sample (3.7$\%$ of their total sample). \citet{Cardamone_2009} imposing conditions on r$_{SDSS}$, redshift, optical colours, and morphology reported a spatial density of Green Peas of approximately 2 per square degree, a critical benchmark in EELG studies. In the local Universe, \citet{2017ApJ...844..171Y} found a density of 0.003 EELGs per square degree for 'blueberries' in the local universe (z $\leq$ 0.05), while \citet{Lumbreras-Calle_2022} identified 466 EELGs within 2000 square degrees, equivalent to a density of about 0.23 per square degree, in the J-PLUS survey (z $\leq$ 0.06). Furthermore, \citet{Amorin_2015} reported 165 EELGs in the 1.7 square degree area covered by the zCOSMOS-bright survey, indicating a density of approximately 97 per square degree in the redshift
range 0.11 $\leq$ z $\leq$ 0.93. As \citet{Jorge_2022_EELG_JPAS} noted precise comparisons among different works  are prevented by the different selection criteria and observational limits of each sample.

There is not a clear limit for the rest-frame equivalent width (EWo) to define a galaxy as an EELG. In this work, we use a similar definition for EELGs to that used by \cite{Jorge_2022_EELG_JPAS}, i.e. objects that show at least one emission line ([OII]$\lambda\lambda$3726,3729, [OIII]$\lambda\lambda$4959,5007 or H$\alpha$) with EWo $\geq$ 300\,\AA{}.  

Nevertheless, we are aware that previous works have assumed different selection limits for EELGs (e.g. \citealt{Amorin_2015,Salzer_2020,Jorge_2022_EELG_JPAS} and references therein); therefore for the sake of consistency with previous works, we have enlarged our search producing an extended sample including all sources presenting EWo $\geq$200\, \AA. 
In this way, we gain consistency in order to compare with literature work, and could examine the behaviour of the main properties of EELGs around the adopted EWo limit; an exercise especially relevant when looking for any possible evolution through redshift.

In this work we take advantage of the exceptional depth and quality of the recent MUSE Hubble Ultra-Deep field survey data release (HUDFS; \citealp{MUSE_DR2_2023}) to carry out a purely spectroscopic search for EELG with MUSE. EELGs usually show a compact appearance although some low surface brightness structure can be seen for some of them (see Appendix A from \citealt{Jorge_2022_EELG_JPAS} for examples). Here, 
we have been able to spatially resolve the ionization structure and the gaseous kinematics of this type of objects.   
This paper is organised as follows. In the next section 
 we summarise the observations and criteria to select the  sample. Section \ref{Sec:Properties} shows the basic properties of our EELG candidates. The results and discussion are presented in Section \ref{sec:results}. The main conclusions are summarised in Section \ref{Sec:Conclusions}.

\section{Data and sample}\label{Sec:Data_sample}

\subsection{MUSE Hubble Ultra-Deep Field}\label{Sec:Data}

The MUSE instrument is a powerful tool for optical (4750–9350\,\AA{}) spectroscopic studies thanks to a combination of a wide field of view (FoV) of 1'x1' with a spatial sampling of 0.2''x 0.2'' per spaxel and high resolving power (ranging from R\,$\sim$\,1800 in the blue to R\,$\sim$\,4000\,\AA{} in the red). Our study is based in the second data release (DR2) of the MUSE Hubble Ultra-Deep Field surveys \citep{MUSE_DR2_2023}, which includes the deepest spectroscopic survey ever performed. With an achieved depth of 141 hours and a field of view of 1 arcmin of diameter, it is a key addition to the previous MUSE spectroscopic survey data release (\citealt{MUSE_DR1_2017, Inami_DR1_2017}; DR1, a 3 × 3 arcmin$^{2}$ mosaic of nine MUSE fields at a 10-hour depth and a single 1 × 1 arcmin$^{2}$ at 31-hour depth) in the Hubble ultra-deep field (\citealp{HUDF_2006}; HUDF) area. This second data release (DR2) incorporates the previous ones reprocessed with the same tools and methodology, providing a homogeneous data set of deep spectroscopic observations in the HUDF region.

In addition to the calibrated data, the data cubes have an extension with the propagated variance estimate, which is used in the line fitting process (see
Sect. \ref{Sec:line_fluxes}). Reduced and calibrated data cubes, the source catalogues and an interactive browser are available to the public through the AMUSED web interface\footnote{\url{https://amused.univ-lyon1.fr/}}.

\subsection{Sample selection}

Our selection process begins by analysing the information offered by the AMUSED interface. The main catalogue gives a summary of source properties and the final datacubes and associated images (e.g. Hubble images) are public for each object. We follow the criteria in \cite{Jorge_2022_EELG_JPAS} and define our primary sample of EELG as those objects that show  at least one emission line with EWo $\geq$ 300\,\AA{}. We use the EWo measured by   \citet{MUSE_DR2_2023} using the python code pyPlatefit\footnote{\url{https://pyplatefit.readthedocs.io/en/latest/tutorial.html}}. Briefly, pyPlatefit is an improved Python-based version of the PLATEFIT IDL software \citep{Brinchmann_2004} developed to fit MUSE data. It performs a stellar continuum fitting using a simple population model \citep{Bruzual_Charlot_2003} and then fits the emission lines with gaussian profiles after subtracting the continuum. To characterise the local continuum, \texttt{pyPlatefit} masks the emission lines and uses windows of 50\,\AA{} on the left and right.

Using these criteria, of the 2221 catalogued objects, we found a total of 2, 5 and 3 galaxies presenting rest-frame EWo $\geq$ 300\,\AA{} in [OII], [OIII] and H$\alpha$, respectively. One of these objects presents EWo $\geq$ 300\,\AA{} in two lines ([OIII] and H$\alpha$). Thus, we selected 9 candidates (ID 91, 891, 895, 1795, 2478, 2532, 6465, 7373 and 7601) fulfilling the criteria for our primary EELG sample. A detailed closer inspection of the spectra of these 9 candidates has been performed on a one by one basis. In doing so, we have found that the spectra of two of the candidates (ID 1795 and 7601) show low signal to noise ratio (SNR) over the entire wavelength range, giving non realistic lines fitting in the AMUSED database. Therefore, these two candidates were discarded from our EELGs primary sample. Table \ref{tab:basic_properties} shows the basic properties of the sample; these include quantities extracted from AMUSED (MUSE ID, coordinates, redshift, magnitude in the HST F775W band, stellar mass, star formation rate, line of detection, line flux, corresponding luminosity\footnote{Note that this luminosity is not corrected by the extinction of each galaxy. See Sect. \ref{sec:phy_props}} and rest frame EWo), and measurements of this work for line flux and rest frame EWo. All these relevant spectrophotometric properties will be considered in the discussion.

Furthermore, as commented before, we have produced also an extended EELG sample with sources presenting 200 \AA{} $\leq$ EWo $\leq$ 300 \AA{} in at least one of the lines considered above ([OII]$\lambda\lambda$3726,3729, [OIII]$\lambda\lambda$4959,5007 or H$\alpha$). Using these extended criteria, we found a total of 3, 8 and 1 galaxies showing 200\,\AA{} $\leq$ EWo $\leq$ 300\,\AA{} in [OII], [OIII] and H$\alpha$, respectively. One of the objects detected in [OIII] presents a EWo $\geq$ 300\,\AA{} in H$\alpha$ (ID 7373) and one object shows 200\,\AA{} $\leq$ EWo $\leq$ 300\,\AA{} in both [OIII] and H$\alpha$ (ID 6474). Therefore, giving 10 candidates (ID 1093, 1426, 1561, 1699, 6474, 6865, 1863, 7105, 7985, 8000) for the extended EELG sample. We have performed a detailed closer inspection of the spectra of these 10 candidates, as done for the primary sample. Following the same criteria, we end up discarding 4 galaxies (ID 1863, 7105, 7985 and 8000). The galaxies selected for our extended EELG sample are listed in Table \ref{tab:basic_properties_extended_sample}. 

\begin{figure*}
\begin{center} 

\includegraphics[width=\textwidth]{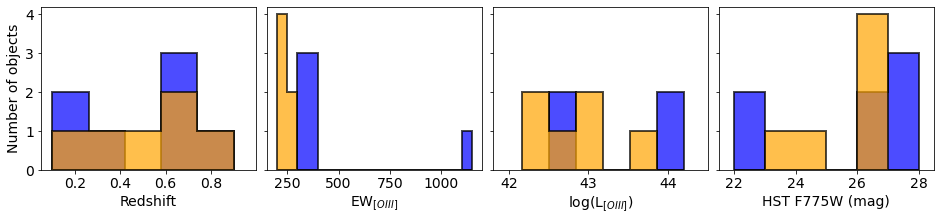}

\caption[]{Histograms of the distribution  of properties of the sample of EELGs. See the text for details.} \label{fig:histograms_sample} 
 \end{center}
\end{figure*}

Figure \ref{fig:histograms_sample} shows the distribution of redshift, mag$_{F775W}$, EWo$_{[OIII]}$\footnote{Note that two galaxies (MUSE ID: 891 and 7373) are included in the primary sample performing its condition in EWo$_{H\alpha}$, though their EWo$_{[OIII]}$ are lower than 300 \AA .} and L$_{[OIII]}$. These candidates expand the ranges: 0.1 $<$ z $<$  0.9, 22 $<$ mag$_{F775W}$ $<$  28, 200\,\AA{} $<$ EWo $<$  1200\,\AA{} and  41.9 $\leq$ Log(L$_{[OIII]}$/$\frac{erg}{s}$) $\leq$ 44.2, respectively. We can see that both, primary and extended, samples share the same range of basic properties. Note that the low number of objects in our sample is totally expected based on reported densities of similar objects in the literature (e.g. \citealt{Jorge_2022_EELG_JPAS}). Furhtermore, we do not find systems with redshift $\leq$0.1 being consistent with previous works in the local Universe (e.g. \citealt{Yang_2017, Lumbreras-Calle_2022}).

\section{Properties of the EELG sample}\label{Sec:Properties}

\subsection{EELG spectra and line fluxes}\label{Sec:line_fluxes}

\begin{figure*}
\begin{center} 

\includegraphics[width=\textwidth]{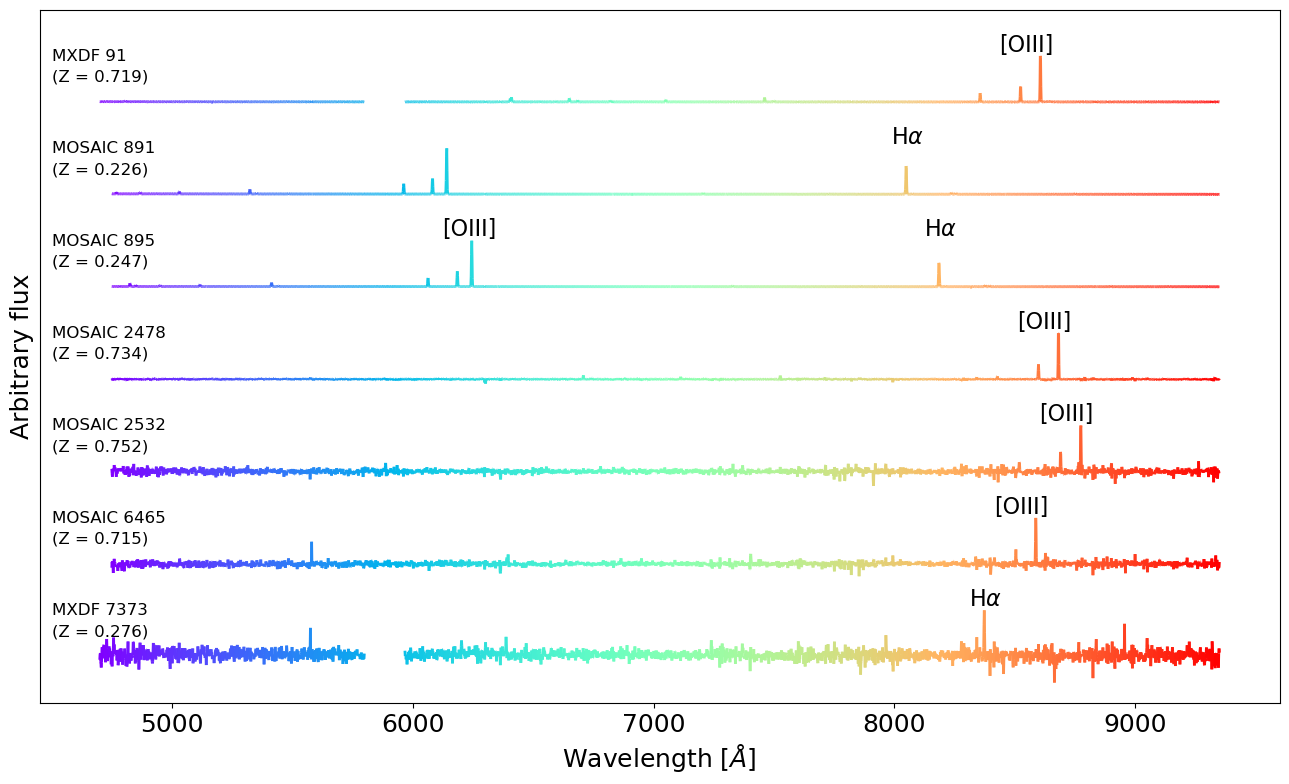}
\includegraphics[width=\textwidth]{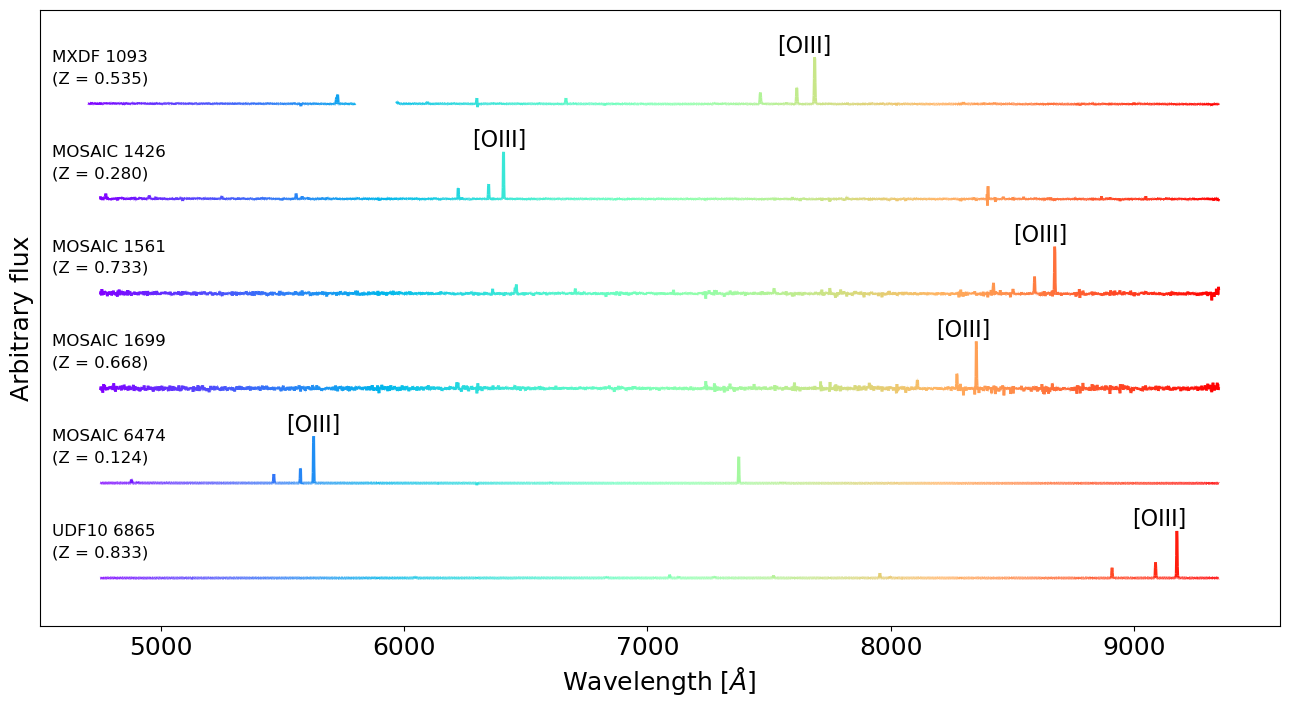}

\caption[]{Spectra for both our primary (\textit{top panel}) and extended (\textit{bottom panel}) EELG samples. MUSE ID, data set, redshift and emission line used for the classification are indicated for each spectra.} \label{fig:spectra_EELGs_extended} 
 \end{center}
\end{figure*}

\begin{figure*}
\begin{center} 

\includegraphics[width=\textwidth]{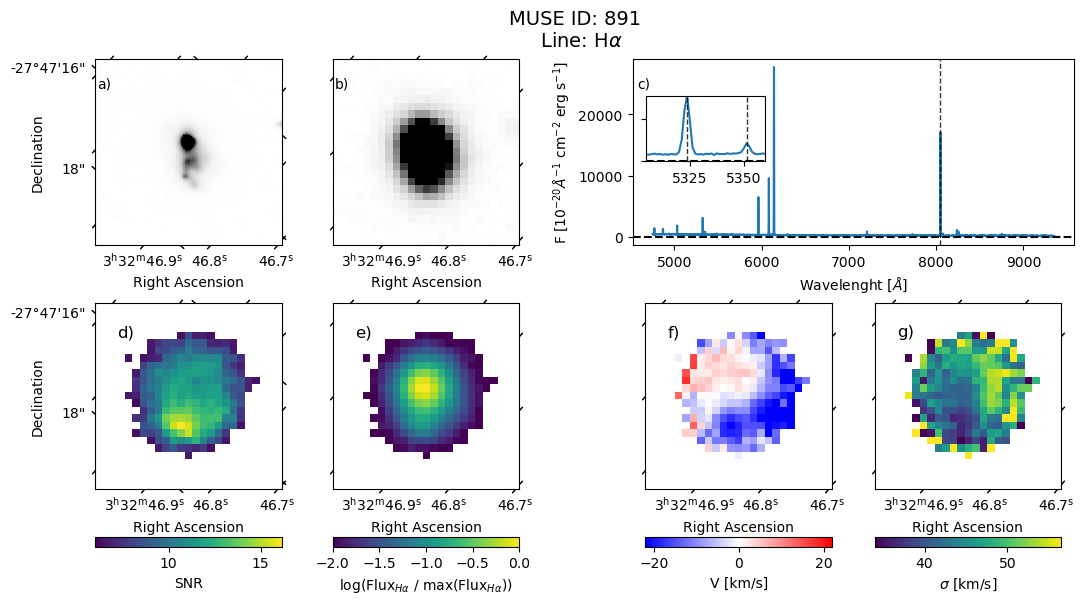}

\caption[]{MUSE ID 891: a) HST F775W image with the spatial coverage of the MUSE data cube; b) MUSE \textit{white light} image of the galaxy; c) spectrum of the object with an inset window showing a zoom around the H$\gamma$ and $[$OIII$]\lambda$4363 emission lines; maps of d) SNR, e) \textbf{normalised} flux, f) relative radial velocity, and g) velocity dispersion ($\sigma$), for the emission line  indicated in the figure header.}

\label{fig:HST_white-light_spectrum_maps_891} 
 \end{center}
\end{figure*}

\begin{table*}
\begin{center}
\resizebox{18.5cm}{!} {
\begin{tabular}{ c c c c c c c c c c c  c | c c c c c c c c}
\hline
\hline
Muse ID  & Ra     & Dec    & z & mag$_{F775W}$  &  log Mass & log SFR & Line & F  & Log$L$ & Cont & EWo & F & Cont & EWo \\
  &    (AMUSED)  &  (AMUSED)   & (AMUSED) & (AMUSED) & (AMUSED) & (AMUSED) & & (AMUSED) & (AMUSED) & (AMUSED) & (AMUSED) & (this work) & (this work) & (this work) \\
         &          (deg)  & (deg)  &   &      & log($\frac{M_{star}}{M\odot}$) &  log($\frac{M\odot}{yr}$) & & (10$^{-20}\frac{erg}{s cm^{2}}$) &($\frac{erg}{s}$) &  (10$^{-20}\frac{erg}{s cm^{2}  \AA{}}$)  & (\AA{}) & (10$^{-20}\frac{erg}{s cm^{2}}$) & (10$^{-20}\frac{erg}{s cm^{2} \AA{}}$) & (\AA{}) \\
   (1)      &    (2)      & (3)  & (4)  & (5)  &  (6)    &  (7) & (8) & (9) & (10) & (11) & (12) & (13) & (14) & (15) \\
\hline
91   & 53.1625 & -27.7803 & 0.72 & 26.18 & 8.26 & -1.27 & [OIII]$\lambda$5007 & 4098  & 42.82 & 11.89  & 345        & 4099$\pm$6   & 11.89$\pm$0.45    & 345$\pm$13 \\
891  & 53.1957 & -27.7878 & 0.23 & 22.21 & 8.44 & -0.54 & H$\alpha$           & 67946 & 44.04 & 218.37 & 311        & 67942$\pm$15 & 218.37$\pm$0.46   & 311$\pm$1 \\
895  & 53.1447 & -27.7854 & 0.25 & 22.47 & 8.6  & -0.64 & [OIII]$\lambda$5007 & 92674 & 44.17 & 255.45 & 363        & 92673$\pm$17 & 255.45$\pm$1.02   & 363$\pm$2 \\
     &         &          &      &       &      &       & H$\alpha$           & 57498 & 43.96 & 173.54 & 331        & 57500$\pm$9  & 173.54 $\pm$ 0.46 & 331$\pm$1 \\
2478 & 53.1839 & -27.7954 & 0.73 & 27.85 & 7.19 & -2.11 & [OIII]$\lambda$4959 & 583   & 41.97 & 1.55   & 375        & 583$\pm$3    & 1.55$\pm$0.26     & 375$\pm$98 \\
     &         &          &      &       &      &       & [OIII]$\lambda$5007 & 2065  & 42.52 & 0.18   & 11478$^\&$ & 2065 $\pm$ 3 & 1.55$\pm$0.26     & 1332$\pm$224 \\
2532 & 53.1497 & -27.8093 & 0.75 & 27.91 & 7.44 & -1.63 & [OIII]$\lambda$5007 & 492   & 41.9  & 1.27   & 387        & 492$\pm$5    & 1.27$\pm$0.37     & 386$\pm$97 \\
6465 & 53.1942 & -27.7854 & 0.72 & 26.15 &  -   & -     & [OIII]$\lambda$5007 & 668   & 42.03 & 1.85   & 362        & 668$\pm$6    & 1.85$\pm$0.46     & 362$\pm$89 \\
7373 & 53.1542 & -27.7867 & 0.28 & 27.89 & 6.85 & -2.43 & H$\alpha$           & 102   & 41.21 & 0.33   & 307$^\&$   & 102$\pm$2    & 0.33$\pm$0.11     & 309$\pm$100 \\

\hline
\hline
\end{tabular}
}
\caption[]{Basic properties of the EELG primary sample candidates. (1) MUSE source identifier;  (2) Right ascension (J2000.0); (3) Declination (J2000.0); (4) Redshift; (5) Magnitude in HST F775W; (6) Stellar mass; (7) Star formation rate; (8) Detected emission line with $\geq$ 300\,\AA{}; (9) Flux of the emission feature; (10) Luminosity of the emission feature; (11) Flux of the continuum level; (12) Rest frame equivalent width of the emission feature ($^\&$ uncertain value); (13) Flux of the emission feature (measured in this work); (14) Flux of the continuum level (measured in this work); (15) Rest frame equivalent width of the emission feature (measured in this work).}

\label{tab:basic_properties} 
\end{center}
\end{table*}

\begin{table*}
\begin{center}
\resizebox{18.5cm}{!} {
\begin{tabular}{ c c c c c c c c c c c  c | c c c c c c c}
\hline
\hline
Muse ID  & Ra     & Dec    & z & mag$_{F775W}$  &  log Mass & log SFR & Line & F  & Log$L$ & Cont & EWo & F & Cont & EWo \\
  &    (AMUSED)  &  (AMUSED)   & (AMUSED) & (AMUSED) & (AMUSED) & (AMUSED) & & (AMUSED) & (AMUSED) & (AMUSED) & (AMUSED) & (this work) & (this work) & (this work) \\
         &          (deg)  & (deg)  &   &      & log($\frac{M_{star}}{M\odot}$) &  log($\frac{M\odot}{yr}$) & & (10$^{-20}\frac{erg}{s cm^{2}}$) &($\frac{erg}{s}$) &  (10$^{-20}\frac{erg}{s cm^{2}  \AA{}}$)  & (\AA{}) & (10$^{-20}\frac{erg}{s cm^{2}}$) & (10$^{-20}\frac{erg}{s cm^{2} \AA{}}$) & (\AA{}) \\
   (1)      &    (2)      & (3)  & (4)  & (5)  &  (6)    &  (7) & (8) & (9) & (10) & (11) & (12) & (13) & (14) & (15) \\
\hline
1093 & 53.1763 & -27.7809 & 0.54 & 24.87 & 7.92 & -0.98 & [OIII]$\lambda$5007 & 5371  & 42.93 & 22.79 & 236 & 5364$\pm$9 & 22.79$\pm$0.38   & 235$\pm$4 \\
1426 & 53.1473 & -27.8008 & 0.28 & 26.05 & 7.07 & -2.06 & [OIII]$\lambda$5007 & 2485  & 42.6  & 8.51 & 292 & 2485$\pm$5  & 8.51$\pm$0.29   & 292$\pm$10 \\
1561 & 53.153  & -27.7937 & 0.73          & 26.41 & 8.1  & -1.3  & [OIII]$\lambda$5007 & 1450  & 42.37 & 5.72 & 254 & 1450$\pm$6  & 5.72 $\pm$0.40   & 254$\pm$18 \\
1699 & 53.154  & -27.8052 & 0.67 & 26.70 & 7.93 & -1.26 & [OIII]$\lambda$5007 & 905   & 42.16 & 3.98 & 228 & 905$\pm$5   & 3.98$\pm$0.24   & 228$\pm$14 \\
6474 & 53.1866 & -27.7902 & 0.12          & 23.20 &   -  &   -   & [OIII]$\lambda$5007 & 26044 & 43.62 & 105.11 & 248 & 26044$\pm$8 & 105.11$\pm$0.39   & 248$\pm$1 \\
     &         &          &               &       &      &       & H$\alpha$  & 15343 & 43.39 & 73.78 & 208 & 15343$\pm$7 & 73.78$\pm$0.48   & 208$\pm$1 \\
6865 & 53.1604 & -27.7752 & 0.83          & 26.60 &   -  &   -   & [OIII]$\lambda$5007 & 6276  & 43.00 & 25.73 & 244 & 6276$\pm$3  & 25.73$\pm$0.24   & 243$\pm$2 \\

\hline
\hline

\end{tabular}
}
\caption[]{Same as Table \ref{tab:basic_properties} for the EELG extended sample candidates.

}
\label{tab:basic_properties_extended_sample} 
\end{center}
\end{table*}




\begin{figure*}
\begin{center}

\includegraphics[width=\textwidth]
{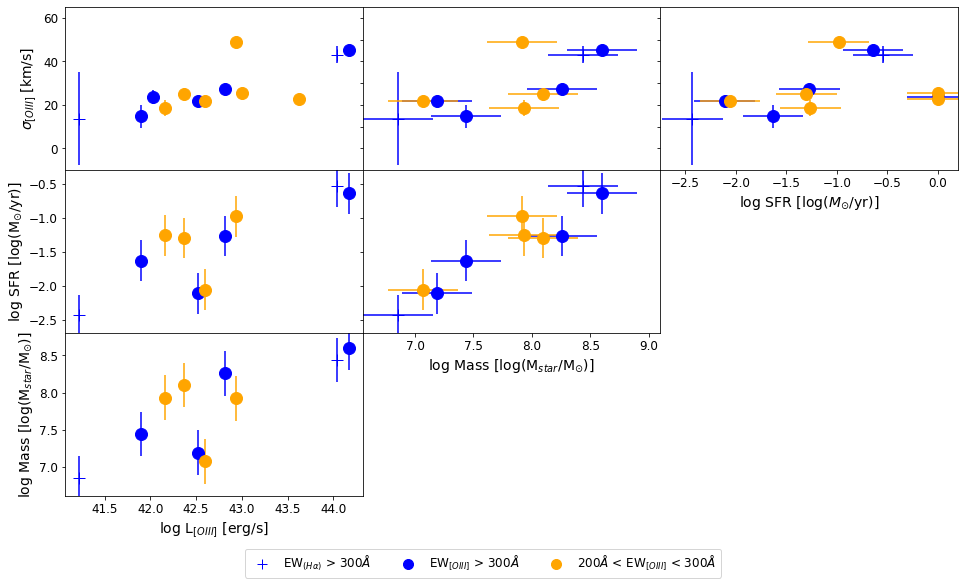}

\caption[]{Correlations between the luminosity, stellar mass, star formation rate and $\sigma_{[OIII]}$. We assume typical uncertainties of 0.3dex for mass and star formation rate (see \citealt{MUSE_DR2_2023}).} \label{fig:figures_properties_EELGS-1} 
 \end{center}
\end{figure*}

Figure \ref{fig:spectra_EELGs_extended} shows the spectra of both EELG primary and extended samples. All the selected objects present characteristic spectra of strong emission line galaxies, consisting of a faint continuum and very prominent emission lines.

As we indicated in Sec. \ref{Sec:Data_sample}, our sample selection is based in the information given by the AMUSED web interface and the fittings performed by \cite{MUSE_DR2_2023}. The equivalent width values depend on the  continuum level and this can be difficult to characterise. Therefore,  we decide to re-evaluate the EWo values of our samples to confirm the EELG classification. Emission-line fitting was performed with \texttt{pyPlatefit}  including improvements with respect the previous fittings. Firstly, we use the parameter \texttt{ziter}, which allow to a second iteration improving the fitting. Secondly, we perform  100 Monte Carlo simulations to compute uncertainties following standard procedures in full spectra fitting techniques (e.g. \citealt{delMoral2019,delMoral2020, Bittner_2019, Jesus_2023Natur} and references there). Briefly, we derive these uncertainties creating multiple realisations of the original spectrum adding random noise consistent with the quality of the observed spectrum to the best fit and running \texttt{pyPlatefit} again. The considered errors are the standard deviation of the 100 recovered values. Tables \ref{tab:table_fluxes_primary} and \ref{tab:table_fluxes_extended} presents the fluxes measured for the emission lines for the primary and extended samples.

Thirdly, three objects (MUSE ID 91, 7373 and 1093) present a cut in the spectrum (from $\sim$ 5800 to  $\sim$5970\,\AA{}). Then, to avoid possible problems with the fitting to the continuum, we fit the wavelength range from 6000 to 9300\,\AA{}. As for the rest of objects, the whole wavelength range was considered (4750-9300\,\AA{}). We notice that the object 2478 presents a very high EWo([OIII]{$\lambda$5007). This value seem to be too high, since the F([OIII]$\lambda$4959) / ([OIII]$\lambda$5007) is $\sim$ 1/3. Therefore, EWo([OIII]$\lambda$5007) should be $\sim$ 3$\cdot$EWo([OIII]$\lambda$4959). However, it is $\sim$30. Through an inspection of the line catalogue given by AMUSED, we found that the continnum associated to the [OIII]$\lambda$5007 emission line is almost an order of magnitude less than the continnum level of the [OIII]$\lambda$4959 emission line. However, F$_{cont}$([OIII]$\lambda$4959) / F$_{cont}$([OIII]$\lambda$5007) $\sim$ 1 for the rest of the objects. Therefore, we decide using the continuum level of the [OIII]$\lambda$4959 emission line to compute the EWo([OIII]$\lambda$5007) of MUSE ID 2478. 

Finally, we studied the effect of the window width used by \texttt{pyPlatefit} to characterise the local continuum\footnote{Although, the broad-band Hubble Space Telescope (HST) observations are deeper than those of MUSE, the filter widths (F606W: 4632.15-7179.43\AA{} and F775W: 6801.41-8630.74\AA{}, see Fig. 1 from \citealt{MUSE_DR2_2023}) are too large to ignore the presence of non-uniform spectral energy distribution (SED) components and other emission features in the characterisation of the local continuum.} on the EWo values. We used \texttt{pyPlatefit} changing the internal configuration to use windows of 100 and 200\,\AA{} confirming the EELG classification except for the object MUSE ID 7373, where the EWo values for windows of 100 and 200\,\AA{} are 102$\pm$11 and 106$\pm$7, respectively. Therefore, we added a flag of caution for this galaxy in Table \ref{tab:basic_properties}.

\subsection{Spectroscopic and spatially resolved properties}\label{Sec:spatially_properties}

Figure \ref{fig:HST_white-light_spectrum_maps_891} presents an example of photometric and spectroscopic properties derived for the galaxy ID 891. The upper row shows the HST 775W  and MUSE white-light images, as well as the MUSE spectrum with an inset zooming around the  H$\gamma$ and $[$OIII$]4363$ \AA \ emission lines. In the bottom row we present the spatially resolved spectroscopic and kinematic information of the galaxy which will be discussed in \ref{Sec:velocitymaps}. Emission-line fitting was performed using pyPlatefit spaxel by spaxel.  
The velocity field shows a clear pattern of rotation similar to those of nearby disc-galaxies (i.e. \citealt{Barrera-Ballesteros2014, delMoral2019}). Only spaxels with significant line detections (S/N $>$ 6) are shown. We checked that the results are not contingent upon the choice of the minimum signal to noise ratio S/N by repeating the analysis with different thresholds (3, 6, and 9). We prefer to stick to S/N=6 to preserve the best possible spatial resolution for this analysis combining with a robust selection. Appendices \ref{sec:App_EELGs_HST_images} and \ref{sec:App_extended_HST_images} show similar images for the  the EELG primary and extended samples, respectively.

\section{Results and discussion}\label{sec:results}

\subsection{Physical properties and chemical abundances}\label{sec:phy_props}

In order to compute the physical properties of the galaxies of our EELG sample\footnote{A closer inspection of all individual line profiles confirmed the presence of narrow lines in all the sources, discarding AGN contamination. Furthermore, the BPT diagram was performed for three of them locating the galaxies in the star formation region.} we have followed the expressions in \cite{EPM_2017PASP}. The Balmer decrement was used for the derivation of the reddening coefficient C(H$\beta$) assuming Case B approximation for electron temperature 15000 K and density 100 cm$^{-3}$ \citep{StoreyHummer1995}, and the extinction law by \cite{Cardelli_1989} with R$_V$=3.1. In any case, no relevant change is expected if the exact electron temperature and density of each object are used. Whenever possible, the ratio of the bright Balmer lines H$\alpha$/H$\beta$ has been used. When the flux measurement of H$\alpha$ (ID  1426) or H$\beta$ (ID 6465) is uncertain, or H$\alpha$ is not observed (ID 2478). The C(H$\beta$) values for all the EELGs are presented in Table \ref{tab:chemical_abundances}. After inspection of the errors of the flux of the Balmer lines considered above (see Appendices \ref{sec:App_fluxes_EELG} and \ref{sec:App_fluxes_extended}), a  formal 20$\%$ relative error, face value, has been adopted for C(H$\beta$).

For two galaxies (ID 2478, ID 6465) the computed C(H$\beta$) was zero or slightly negative, though consistent to within the errors with the adopted value of C(H$\beta$)=0. A C(H$\beta$)=0 was adopted also for ID 2532 since only one Balmer line, H$\beta$, was measured for this galaxy. In what follows, all the fluxes used for the derivation of the physical properties and chemical abundances of the selected EELGs have been reddening corrected accordingly. 

Overall, the extinction values derived for our sample of EELGs are low, typically C(H$\beta$) $\approx$ 0.1 (Av $\approx$ 0.22), and the  maximum around Av $\approx$ 0.45, whereas a good fraction of the sample  shows values consistent with no extinction. This result, -given the negligible Galactic extinction,  \citep{Extinct_NED2011}-  would translate into a low (or very modest) dust component associated to our EELGs, somewhat expected for low metallicity (dwarf) galaxies.

For seven galaxies the flux in the auroral [OIII]$\lambda$4363 temperature sensitive line has been measured with good signal to noise, illustrating the high quality of the data. Therefore we have performed a direct calculation of their electron temperature, with important implications for the chemical abundance derivation. We have followed \cite{EPM_2017PASP} for the computation of the T[OIII] electron temperature using the ([OIII]$\lambda$5007 +[OIII]$\lambda$4959)/[OIII]$\lambda$4363 line ratio. The values of T[OIII] obtained, along with their corresponding errors derived via error propagation from the measured flux error, are presented in 
Table \ref{tab:chemical_abundances}. The T[OIII] electron temperatures are high, going from 1.3 10$^4$ K to over 20 10$^4$ K; these temperature values being typically found in low metallicity star-forming galaxies (e.g. \citealt{Amorin_2015, Kehrig_2016, Izotov_2019, Perez_montero_2021}).

The electron density, N$_e$, has been derived using the line ratios from the [OII]$\lambda \lambda$3726,3729 and [SII]$\lambda \lambda$6717,6731 doublets whenever observed (redshift permitting, see Fig. \ref{fig:inset} for an example), being both line doublets resolved in the MUSE data. For galaxies ID 91, 6465, 1093, 1426, 1561, 1699, 6865, we have measurements of the [OII]$\lambda \lambda$3726,3729 doublet; whereas the [SII]$\lambda \lambda$6717,6731 doublet is measured for ID 891, 895, 6474. For three galaxies, ID 2478, 2532, 7373, we have no information to derive their electron density. 
We assume a electron temperature of 10$^4$ K
\citep{Sanders_2016, Harshan_2020}. After a close inspection of the measured flux errors, in this work we have adopted a formal relative error for N$_e$ of 20$\%$, face value, and value of the low density limit for N$_e$$\leq$ 50 cm$^{-3}$. The electron densities derived for our EELG  galaxies are shown in Table \ref{tab:chemical_abundances}. These values of electron density indicate that most galaxies of the sample have a low density, and N$_e$ in four of them being consistent with the low density limit. However, for three galaxies, ID 1699, 6474, 6865, substantially high electron densities are derived, with N$_e$$\approx$ 500 to 600 cm$^{-3}$. Interestingly, two of them, ID 6474 and ID 6865, show the highest electron temperatures derived. 

\begin{figure}
\begin{center} 

\includegraphics[width=0.24\textwidth]{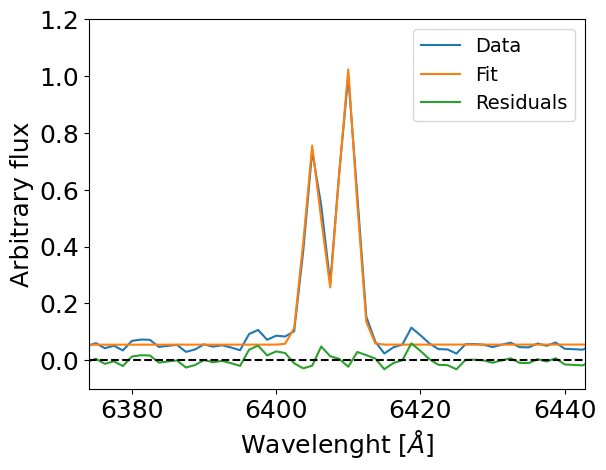}
\includegraphics[width=0.24\textwidth]{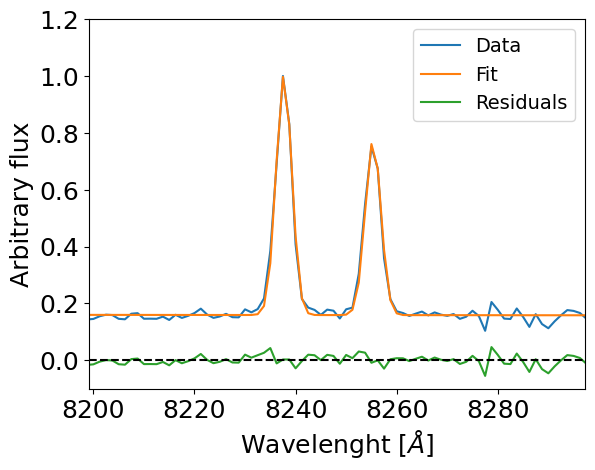}

\caption[]{Electron density sensitive lines. \textit{Left panel:} fit to the  $[$OII$]\lambda\lambda$3726,3729 doublet in the spectrum of galaxy ID 91. \textit{Right panel:} fit to the  $[$SII$]\lambda\lambda$6717,6731 doublet in the spectrum of galaxy ID 891. In each plot, the observed spectrum (blue) and its fit (orange), and the residuals (green), are shown.} \label{fig:inset} 
 \end{center}
\end{figure}

After obtaining the physical properties of the galaxies above we have derived their metallicity using the spectroscopic information gathered. We have calculated the oxygen chemical abundance of the ionized gas of these EELGs, as a direct measurement of their gas metallicity. To do so we proceed as follows: 

\begin{enumerate}
    \item For those galaxies (ID 91, 1093\footnote{A second fitting in the rest-frame 4000-5800\AA{} was perfomed to fit the [OII]$\lambda \lambda$3726,3729 doublet.}, 6865) for which we have measurements of [OII]$\lambda \lambda$3726,3729 and [OIII]$\lambda \lambda$5007,4959 flux, together with the electron temperature T[OIII], we can perform a direct derivation of the ionic O$^+$/H$^+$ and O$^{++}$/H$^+$ abundances, adding them up to derive the total oxygen abundance, expressed  12+log(O/H). For this calculation a two ionization zones scheme has been adopted following \cite{EPM_2017PASP}, using T[OIII] for the O$^{++}$ zone, and estimating the O$^+$ zone temperature, T[OII]. The 12+log(O/H) of these galaxies, along with their corresponding errors estimated from error propagation of line fluxes and temperature, are quote in Table \ref{tab:chemical_abundances}.

    \item For those galaxies (ID 891, 895, 2478, 6474) for which we have derived the electron temperature, T[OIII], but no [OII]$\lambda \lambda$3726,3729 lines are measured, we can not proceed with the direct method as above. In these cases we rely on the anti-correlation between the total oxygen abundance of the ionized gas and its electron temperature, as observed in \cite{Amorin_2015}; a direct, theoretically expected, consequence of the cooling of the ionized gas via oxygen lines. The values of 12+log(O/H) for these galaxies, along with the corresponding statistical error $\pm$0.08 dex, have been derived following the expression in \cite{Amorin_2015}, and are presented in Table \ref{tab:chemical_abundances}. For the sake of comparison, we have derived 12+log(O/H) also for the three galaxies treated with the direct method above, giving a very good agreement within the errors. 

    \item For the rest of the galaxies we only have flux measurements of  [OII]$\lambda \lambda$3726,3729 and  [OIII]$\lambda \lambda$5007,4959 lines, for ID 6465, 1426, 1561, 1699, or even only [OIII]$\lambda \lambda$5007,4959 for ID 2532, 7373. In this third case, in order to determine the oxygen abundance we have to rely on an empirical calibration of abundance based on bright lines. Among the empirical calibrations, the one based on the R$_{23}$ parameter, originally defined by \cite{Pagel_1979} as R$_{23}$ = ([OII]$\lambda \lambda$3726,3729 +[OIII]$\lambda \lambda$5007,4959)/H$\beta$ stands out as a useful and robust tool against possible ionization structure effects (e.g. \citealt{Relano_2010, Kehrig_2016}. In this work we have made use of the R$_{23}$ calibration \citep{Izotov_2019} to derive the oxygen abundance 12+log(O/H). Though, the R$_{23}$ calibration is bi-valuated and additional information is required to choose the appropriate branch, thus we have applied it as in \cite{Williams_2023}, selecting the low metallicity regime taking into account the high [OIII]$\lambda \lambda$5007,4959/[OII]$\lambda \lambda$3726,3729 measured for the EELG galaxies (e.g. \citealt{Sanders_2016}).  The derived 12+log(O/H) abundances and their corresponding statistical errors of $\pm$0.1 dex are presented in Table \ref{tab:chemical_abundances}. We also show the values of R$_{23}$ when available, or logO$_3$=[OIII]$\lambda \lambda$5007,4959)/H$\beta$ when [OII]$\lambda\lambda$3726,3729 lines are not measured.
\end{enumerate}

It has been reported that the ionised interstellar medium in some extreme emission line galaxies could be experiencing density bounded conditions (e.g. \citealt{JaskotOey_2013}). These objects would show extreme [OIII]$\lambda\lambda$5007, 4959/[OII]$\lambda\lambda$3737,29 flux ratios (typically $\gtrsim$ 40) suggestive of the presence of non-standard conditions which may complicate the applicability of abundance calibrations. The abundance calibration applied in this work is selected from \cite{Izotov_2019} and it was specifically derived for the class of extreme emission line low metallicity galaxies studied here. All sample galaxies for which the empirical calibration has been applied present moderate to low values of the [OIII]$\lambda\lambda$5007, 4959/[OII]$\lambda\lambda$3737,29 ratio (see Tables C.1, D1), hence any potential effect associated to density bounded conditions is expected to be minor.
The analysis of the oxygen abundances derived for our EELG sample indicates that these galaxies are all metal poor, with 7.35 $\leq$ 12+log(O/H) $\leq$ 8.05, showing four of them a metallicity below 0.1 solar. The very low metallicities obtained in our search for EELGs have important implications for their characterization as local analogs of the earliest star-forming galaxies.

\begin{table*}
\begin{center}
\begin{tabular}{ c c c c c c c c c c c c c c c c c}
\hline
\hline
ID & C(H$\beta$) & Te($[$OIII$]$) & Ne & logR$_{23}$ & 12+log(O/H) & Notes \\
  &  & (10$^4$\,K)  & (cm$^{-3}$) & (logO$_3$) & &  \\
\hline
91	            & 0.11		& 1.65$\pm$0.04		& $\leq$ 50 & 	0.89		& 7.72$\pm$0.03 [7.75]	& (b) P \\
891	            & 0.17		& 1.45$\pm$0.02		& 73		& (0.76)		& 7.93$\pm$0.08	    & (a) P \\
895	            & 0.18		& 1.38$\pm$0.01		& 75 		& (0.82)		& 7.99$\pm$0.08	    & (a) P\\
2478$^{(\star)}$ & 0.00		& 1.49$\pm$0.20	    & 	 - 		& (0.95)		& 7.89$\pm$0.08	    & (a) P	\\
2532$^{(\&)}$  & 0.00	    & 	 -	    & 	 -	    & 	(0.77)	    & 	   -		& (d) P \\
6465    & 0.00		&    -		& 90 		& 1.03		    & 7.77$\pm$0.10	& (c) P \\
7373	        & 0.22		&    -		&  -		& (0.14)		&    -		    & (d) P \\
1093	        & 0.09		& 1.31$\pm$0.12		& $\leq$ 50	& 	0.81	    & 7.90$\pm$0.10 [8.05]	& (b) E \\
1426            & 0.20		&    -		& $\leq$ 50	& 	0.75	    & 7.47$\pm$0.10   & (c) E \\
1561            & 0.13		&    -		& $\leq$ 50	& 	0.83		& 7.58$\pm$0.10  & (c) E \\
1699 	        & 0.10		&    -		& 544 	    & 	0.77		& 7.51$\pm$0.10   & (c) E \\
6474	        & 0.13		& 1.69$\pm$0.06		& 496 		& (0.83)		& 7.72$\pm$0.08	    & (a) E \\
6865 	        & 0.12		& 2.08$\pm$0.05		& 557 		& 0.76		    & 7.38$\pm$0.02 [7.37]	& (b) E \\
\hline
\hline

\end{tabular}

\caption[]{Physical properties and chemical abundances.

($\star$): H$\beta$ flux measurement uncertain.

(\&): Only H$\beta$ line flux measured; C(H$\beta$)=0 assumed.

Notes (Column 7).- 

(a): Oxygen abundance derived from Te($[$OIII$]$) following \citet{Amorin_2015} relation; expected $\pm$0.08 dex statistical error.

(b): Oxygen abundance from the direct method; the abundance derived using (a) is also given in [ ] for comparison.

(c): Oxygen abundance from R$_{23}$ empirical calibration \citep{Izotov_2019, Williams_2023}; expected $\pm$0.10 dex statistical error.  

(d): No oxygen abundance derived; only $[$OIII$]$ nebular line measured.

P: primary EELG sample; E: extended EELG sample.}
\label{tab:chemical_abundances}
\end{center}
\end{table*}

\subsection{Kinematical properties, structure and velocity maps}\label{Sec:velocitymaps}

By inspecting the kinematics maps presented in Sec. \ref{Sec:spatially_properties}, we can separate our sample in three different classes: 
\begin{enumerate}
    \item \textit{Resolved with kinematic pattern\footnote{Note that this class presents significantly larger velocity amplitudes. }:} galaxies spatially resolved that show a clear pattern of rotation (i.e. presence of a disc, with the velocity increasing outward). About $\sim$43$\%$ of EELGs in the primary sample are in this class:

    \begin{itemize}
        \item \textit{ID 891:} The HST F775W continuum image reflects a non-nuclear shape, with several high-surface brightness knots connected by a diffuse low-surface brightness component. This structure is not shown in the MUSE white light image (given the lower spatial resolution), which present a head-tail shape (i.e. a main bright star-forming clump located at the centre of the image and a low-surface brightness tail). The SNR map shows its highest values in the inner galaxy, though peaking slightly south of the flux maximum likely a consequence of the different line profile shapes and the higher velocity dispersion measured. The ionised gas velocity field presents a clear rotating disk structure, showing higher values in the region corresponding to the knots identified in the HST image. Note that these knots could be reminiscent of the structures seen in tidal dwarf and/or cometary galaxies 
    (e.g. \citealt{Lagos_2016, Roche_2023}). These galaxies are usually considered the result of a tidal process or interactions of disk galaxies. No apparent companion is envisaged in this Figure, however further investigation on its true nature is beyond the scope of this paper. 
   
        \item \textit{ID 895:} The HST image shows a elongated shape, also reflected by the MUSE white light image. The $[$OIII$]$ SNR map shows a nuclear shape while the H$\alpha$ map also presents an elongated shape with the highest values peaking slightly south of the flux maximum. Both $[$OIII$]$ and H$\alpha$ flux maps present a nuclear shape peaking at the centre of the galaxy.  The $[$OIII$]$ and H$\alpha$ velocity field show clear rotation disk structures.
    
        \item \textit{ID 91:} The HST F775W image present a nuclear shape. This shape is also shown in the MUSE white light, SNR and flux maps. The $[$OIII$]$ velocity field show a rotating pattern.
        
    \end{itemize}

    \item \textit{Resolved without kinematic pattern:} galaxies spatially resolved but, unlike the previous category, do not show a rotation pattern. This class represents $\sim$14$\%$ and $\sim$83$\%$ of our EELGs (ID: 2478) and extended samples, respectively (ID: 1093, 1426, 1561, 6474 and 6865). The HST F775W images reflect a nuclear shape, except for the ID 6474 object, which present a elongated shape with a low-surface brightness tail. The nuclear shape is also presented in the MUSE white light, SNR and flux maps for the six galaxies.

    \item \textit{Not resolved:} galaxies whose structure can not be resolved. About $\sim$43$\%$ and $\sim$17$\%$ of both, EELGs primary (ID: 2532, 6465 and 7373) and extended (ID: 1699) samples, respectively, are included in this class.
\end{enumerate}
Note that we applied a similar approach as \citet{Guerou_2017} considering  a galaxy spatially resolved if the galaxy has at least an area of 16 MUSE spaxels (i.e. about 1.5 times the PSF FWHM size of the data cube). We are aware that mergers may mimic the appearance of rotation patterns in observations of finite spatial resolution. Therefore, observations with better spatial resolution (e.g. JWST or MUSE-NFM observations) are necessary to reliably determine the internal structure of these systems.

\subsection{Global properties and evolution}

Figure \ref{fig:figures_properties_EELGS-1} displays the relations among the derived [OIII] luminosity and velocity dispersion, stellar mass and SFR (both parameters taken from the AMUSED database, as derived using the Prospector code\footnote{All the details of the derivation of stellar mass and SFR can be found in \citet{MUSE_DR2_2023}}), for both the primary and extended samples. Inspection of this figure first reveals that, when comparing the behaviour of both samples, no clear differences can be found between them, i.e. the EWo limits assumed to select our EELGs do not seem to introduce a bias in their observed properties. Overall, all four parameters show good correlations, with the SFR vs. stellar mass plot delineating a very clear main sequence for our sample of EELGs. From the ionized gas, the derived [OIII] line luminosity and $\sigma$ correlate, showing both parameters some degree of correlation versus stellar mass and SFR -most notably between stellar mass and [OIII] luminosity-. This is relevant since we are comparing properties, e.g. [OIII] luminosity and stellar mass, which have been independently derived, corresponding to gaseous and stellar components of the EELG galaxies, respectively. 

In Figure \ref{fig:mass_VS_metal} we present the mass-metallicity relation showing average loci corresponding to nearby (SDSS) and high redshift galaxies. Our EELG galaxies\footnote{Only eight of the galaxies in Table \ref{tab:chemical_abundances} have stellar mass values in the AMUSED database}, sampling the low stellar mass (Mass$_{pro}\leq$9.0) and low  metallicity (12+log(O/H)$\leq$8.0) region of the plot, are consistent with the mass–metallicity relation derived for z$\sim$2.2  \citep{Sanders_2016} galaxies. They are also close to the locus defined by very high-redshift objects (z = 7.7-8-5 and z = 6.1-6.3), as recently studied by \citet{Schaerer_2022} and \citet{Sun_2023} using JWST data. This result illustrates how the EELGs in our sample can be considered true local analogs of these high-redshift systems. The next generation of ground-based 30-meter telescopes, together with additional observations from space, will allow us to fill this figure with low metallicity and low massive EELG galaxies at high-redshift and their local analogs.

\begin{figure}
\begin{center}

\includegraphics[width=0.45\textwidth]{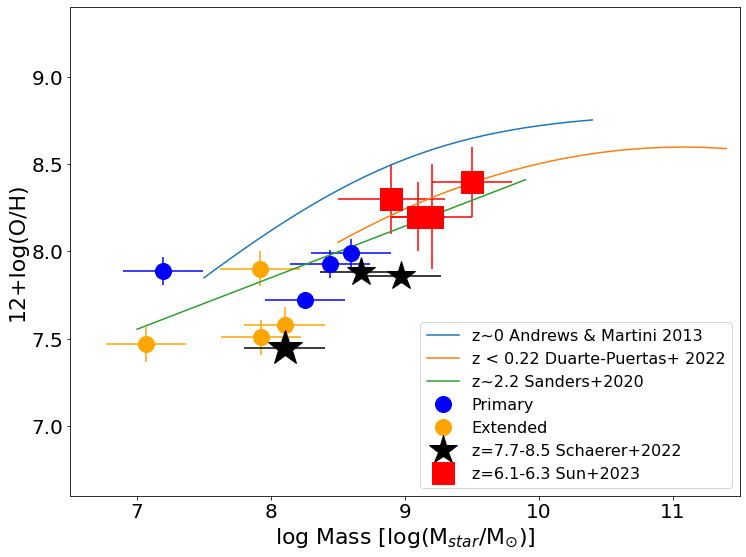}

\caption[]{Mass-metallicity relation. Oxygen chemical abundance and stellar mass for the galaxies in the primary (blue circles) and extended (orange circles) samples in this work, compared to mean relations observed at z$\sim$0 \citep{Andrews_2013}, z$<$0.22 \citep{Duarte_2022}, and  z=2.2 \citep{Sanders_2016}. Black stars and red squares show recent results for high-redshift galaxies from JWST data \citep{Schaerer_2022, Sun_2023}.} \label{fig:mass_VS_metal} 
 \end{center}
\end{figure}

\section{Summary}\label{Sec:Conclusions}

During decades, the EELGs searches have been performed on broad-band photometry data. Recently, thanks to the development of new instruments and recent surveys such as SDSS or J-PAS, direct spectroscopy or narrow-band searches have been carried out (e.g. \citealt{Indahl_2021, Jorge_2022_EELG_JPAS}). However, the number of EELG searches and analyses performed to date based on IFU data is (to our knowledge) very scarce. In fact, this work could be considered among the first EELG searches performed based on deep IFU data surveys, since previous works have studied mostly specific objects (e.g. \citealt{Lofthouse_2017, Bosch_2019, Arroyo-Polonio_2023}).

Taking advantage of the deepest IFU survey to date, we present spatially resolved kinematic maps for a sample of intermediate redshift EELGs observed from ground. Furthermore, we confirm that these systems are true local analogs of high-redshift systems (z=6-8) studied by the JWST. While these cases already reveal the fantastic potential of the MUSE Ultra Deep Field data set, our small sample prevents a more detailed study on the basic properties of EELGs. However, this work can be regarded as a successful pilot study, demonstrating the  power of the presented methodology to analyse strong emission-line galaxies using IFU data. Future deep observations with MUSE or on-going instrument such as HARMONI will result in a larger sample in which to apply the methodology presented here.

\begin{acknowledgements}
We thank the referee for carefully reading our manuscript and providing constructive comments improving the paper. This study uses data provided by the Muse Hubble Ultra Deep survey, publicly available at the AMUSED database\footnote{https://amused.univ-lyon1.fr/}. IMC acknowledges funding from U. La Laguna through the Margarita Salas Fellowship from the Spanish Ministry of Universities ref. UNI/551/2021-May 26 and under the EU Next Generation funds, and from ANID programme FONDECYT Postdoctorado 3230653. JMV, JIP, AAP acknowledge financial support from grants CEX2021-001131-S and PID2019-107408GB-C44 funded by MCIN/AEI/ 10.13039/501100011033. This paper makes use of python\footnote{ http://www.python.org}; Matplotlib \citep{Hunter_2007}: a suite of open-source python modules that provide a framework for creating scientific plots; Astropy\footnote{http://www.astropy.org}: a community-developed core Python package and an ecosystem of tools and resources for astronomy \citep{astropy:2013, astropy:2018, astropy:2022}; MPDAF\footnote{https://mpdaf.readthedocs.io}: an open-source Python package that provides tools to work with MUSE-specific data; and pyplatefit\footnote{ https://github.com/musevlt/pyplatefit} \citep{MUSE_DR2_2023}: a python routine for emission and absorption lines fit of MUSE spectra. and AMUSED\footnote{https://amused.univ-lyon1.fr/}: a public web interface for inspection and retrieval of extragalactic MUSE data products. IMC thank Alejandro S. Borlaff for helpful discussions on the HST broad-band images.

\end{acknowledgements}

%
   \bibliographystyle{aa} 
   \bibliography{biblio} 

\begin{thebibliography}{66}
\expandafter\ifx\csname natexlab\endcsname\relax\def\natexlab#1{#1}\fi

\bibitem[{{Amor{\'\i}n} {et~al.}(2015){Amor{\'\i}n}, {P{\'e}rez-Montero}, {Contini}, {V{\'\i}lchez}, {Bolzonella}, {Tasca}, {Lamareille}, {Zamorani}, {Maier}, {Carollo}, {Kneib}, {Le F{\`e}vre}, {Lilly}, {Mainieri}, {Renzini}, {Scodeggio}, {Bardelli}, {Bongiorno}, {Caputi}, {Cucciati}, {de la Torre}, {de Ravel}, {Franzetti}, {Garilli}, {Iovino}, {Kampczyk}, {Knobel}, {Kova{\v{c}}}, {Le Borgne}, {Le Brun}, {Mignoli}, {Pell{\`o}}, {Peng}, {Presotto}, {Ricciardelli}, {Silverman}, {Tanaka}, {Tresse}, {Vergani}, \& {Zucca}}]{Amorin_2015}
{Amor{\'\i}n}, R., {P{\'e}rez-Montero}, E., {Contini}, T., {et~al.} 2015, \aap, 578, A105

\bibitem[{{Amor{\'\i}n} {et~al.}(2012){Amor{\'\i}n}, {V{\'\i}lchez}, {H{\"a}gele}, {Firpo}, {P{\'e}rez-Montero}, \& {Papaderos}}]{AmorinVilchez2012_rotating}
{Amor{\'\i}n}, R., {V{\'\i}lchez}, J.~M., {H{\"a}gele}, G.~F., {et~al.} 2012, \apjl, 754, L22

\bibitem[{{Amor{\'\i}n} {et~al.}(2010){Amor{\'\i}n}, {P{\'e}rez-Montero}, \& {V{\'\i}lchez}}]{Amorin_2010}
{Amor{\'\i}n}, R.~O., {P{\'e}rez-Montero}, E., \& {V{\'\i}lchez}, J.~M. 2010, \apjl, 715, L128

\bibitem[{{Andrews} \& {Martini}(2013)}]{Andrews_2013}
{Andrews}, B.~H. \& {Martini}, P. 2013, \apj, 765, 140

\bibitem[{{Arroyo-Polonio} {et~al.}(2023){Arroyo-Polonio}, {Iglesias-P{\'a}ramo}, {Kehrig}, {V{\'\i}lchez}, {Amor{\'\i}n}, {Breda}, {P{\'e}rez-Montero}, {P{\'e}rez-D{\'\i}az}, \& {Hayes}}]{Arroyo-Polonio_2023}
{Arroyo-Polonio}, A., {Iglesias-P{\'a}ramo}, J., {Kehrig}, C., {et~al.} 2023, \aap, 677, A114

\bibitem[{{Astropy Collaboration} {et~al.}(2022){Astropy Collaboration}, {Price-Whelan}, {Lim}, {Earl}, {Starkman}, {Bradley}, {Shupe}, {Patil}, {Corrales}, {Brasseur}, {N{"o}the}, {Donath}, {Tollerud}, {Morris}, {Ginsburg}, {Vaher}, {Weaver}, {Tocknell}, {Jamieson}, {van Kerkwijk}, {Robitaille}, {Merry}, {Bachetti}, {G{"u}nther}, {Aldcroft}, {Alvarado-Montes}, {Archibald}, {B{'o}di}, {Bapat}, {Barentsen}, {Baz{'a}n}, {Biswas}, {Boquien}, {Burke}, {Cara}, {Cara}, {Conroy}, {Conseil}, {Craig}, {Cross}, {Cruz}, {D'Eugenio}, {Dencheva}, {Devillepoix}, {Dietrich}, {Eigenbrot}, {Erben}, {Ferreira}, {Foreman-Mackey}, {Fox}, {Freij}, {Garg}, {Geda}, {Glattly}, {Gondhalekar}, {Gordon}, {Grant}, {Greenfield}, {Groener}, {Guest}, {Gurovich}, {Handberg}, {Hart}, {Hatfield-Dodds}, {Homeier}, {Hosseinzadeh}, {Jenness}, {Jones}, {Joseph}, {Kalmbach}, {Karamehmetoglu}, {Ka{l}uszy{'n}ski}, {Kelley}, {Kern}, {Kerzendorf}, {Koch}, {Kulumani}, {Lee}, {Ly}, {Ma}, {MacBride}, {Maljaars}, {Muna}, {Murphy}, {Norman}, {O'Steen},
  {Oman}, {Pacifici}, {Pascual}, {Pascual-Granado}, {Patil}, {Perren}, {Pickering}, {Rastogi}, {Roulston}, {Ryan}, {Rykoff}, {Sabater}, {Sakurikar}, {Salgado}, {Sanghi}, {Saunders}, {Savchenko}, {Schwardt}, {Seifert-Eckert}, {Shih}, {Jain}, {Shukla}, {Sick}, {Simpson}, {Singanamalla}, {Singer}, {Singhal}, {Sinha}, {Sip{H{o}}cz}, {Spitler}, {Stansby}, {Streicher}, {{{S}}umak}, {Swinbank}, {Taranu}, {Tewary}, {Tremblay}, {Val-Borro}, {Van Kooten}, {Vasovi{'c}}, {Verma}, {de Miranda Cardoso}, {Williams}, {Wilson}, {Winkel}, {Wood-Vasey}, {Xue}, {Yoachim}, {Zhang}, {Zonca}, \& {Astropy Project Contributors}}]{astropy:2022}
{Astropy Collaboration}, {Price-Whelan}, A.~M., {Lim}, P.~L., {et~al.} 2022, apj, 935, 167

\bibitem[{{Astropy Collaboration} {et~al.}(2018){Astropy Collaboration}, {Price-Whelan}, {Sip{\H{o}}cz}, {G{\"u}nther}, {Lim}, {Crawford}, {Conseil}, {Shupe}, {Craig}, {Dencheva}, {Ginsburg}, {Vand erPlas}, {Bradley}, {P{\'e}rez-Su{\'a}rez}, {de Val-Borro}, {Aldcroft}, {Cruz}, {Robitaille}, {Tollerud}, {Ardelean}, {Babej}, {Bach}, {Bachetti}, {Bakanov}, {Bamford}, {Barentsen}, {Barmby}, {Baumbach}, {Berry}, {Biscani}, {Boquien}, {Bostroem}, {Bouma}, {Brammer}, {Bray}, {Breytenbach}, {Buddelmeijer}, {Burke}, {Calderone}, {Cano Rodr{\'\i}guez}, {Cara}, {Cardoso}, {Cheedella}, {Copin}, {Corrales}, {Crichton}, {D'Avella}, {Deil}, {Depagne}, {Dietrich}, {Donath}, {Droettboom}, {Earl}, {Erben}, {Fabbro}, {Ferreira}, {Finethy}, {Fox}, {Garrison}, {Gibbons}, {Goldstein}, {Gommers}, {Greco}, {Greenfield}, {Groener}, {Grollier}, {Hagen}, {Hirst}, {Homeier}, {Horton}, {Hosseinzadeh}, {Hu}, {Hunkeler}, {Ivezi{\'c}}, {Jain}, {Jenness}, {Kanarek}, {Kendrew}, {Kern}, {Kerzendorf}, {Khvalko}, {King}, {Kirkby}, {Kulkarni},
  {Kumar}, {Lee}, {Lenz}, {Littlefair}, {Ma}, {Macleod}, {Mastropietro}, {McCully}, {Montagnac}, {Morris}, {Mueller}, {Mumford}, {Muna}, {Murphy}, {Nelson}, {Nguyen}, {Ninan}, {N{\"o}the}, {Ogaz}, {Oh}, {Parejko}, {Parley}, {Pascual}, {Patil}, {Patil}, {Plunkett}, {Prochaska}, {Rastogi}, {Reddy Janga}, {Sabater}, {Sakurikar}, {Seifert}, {Sherbert}, {Sherwood-Taylor}, {Shih}, {Sick}, {Silbiger}, {Singanamalla}, {Singer}, {Sladen}, {Sooley}, {Sornarajah}, {Streicher}, {Teuben}, {Thomas}, {Tremblay}, {Turner}, {Terr{\'o}n}, {van Kerkwijk}, {de la Vega}, {Watkins}, {Weaver}, {Whitmore}, {Woillez}, {Zabalza}, \& {Astropy Contributors}}]{astropy:2018}
{Astropy Collaboration}, {Price-Whelan}, A.~M., {Sip{\H{o}}cz}, B.~M., {et~al.} 2018, \aj, 156, 123

\bibitem[{{Astropy Collaboration} {et~al.}(2013){Astropy Collaboration}, {Robitaille}, {Tollerud}, {Greenfield}, {Droettboom}, {Bray}, {Aldcroft}, {Davis}, {Ginsburg}, {Price-Whelan}, {Kerzendorf}, {Conley}, {Crighton}, {Barbary}, {Muna}, {Ferguson}, {Grollier}, {Parikh}, {Nair}, {Unther}, {Deil}, {Woillez}, {Conseil}, {Kramer}, {Turner}, {Singer}, {Fox}, {Weaver}, {Zabalza}, {Edwards}, {Azalee Bostroem}, {Burke}, {Casey}, {Crawford}, {Dencheva}, {Ely}, {Jenness}, {Labrie}, {Lim}, {Pierfederici}, {Pontzen}, {Ptak}, {Refsdal}, {Servillat}, \& {Streicher}}]{astropy:2013}
{Astropy Collaboration}, {Robitaille}, T.~P., {Tollerud}, E.~J., {et~al.} 2013, \aap, 558, A33

\bibitem[{{Bacon} {et~al.}(2023){Bacon}, {Brinchmann}, {Conseil}, {Maseda}, {Nanayakkara}, {Wendt}, {Bacher}, {Mary}, {Weilbacher}, {Krajnovi{\'c}}, {Boogaard}, {Bouch{\'e}}, {Contini}, {Epinat}, {Feltre}, {Guo}, {Herenz}, {Kollatschny}, {Kusakabe}, {Leclercq}, {Michel-Dansac}, {Pello}, {Richard}, {Roth}, {Salvignol}, {Schaye}, {Steinmetz}, {Tresse}, {Urrutia}, {Verhamme}, {Vitte}, {Wisotzki}, \& {Zoutendijk}}]{MUSE_DR2_2023}
{Bacon}, R., {Brinchmann}, J., {Conseil}, S., {et~al.} 2023, \aap, 670, A4

\bibitem[{{Bacon} {et~al.}(2017){Bacon}, {Conseil}, {Mary}, {Brinchmann}, {Shepherd}, {Akhlaghi}, {Weilbacher}, {Piqueras}, {Wisotzki}, {Lagattuta}, {Epinat}, {Guerou}, {Inami}, {Cantalupo}, {Courbot}, {Contini}, {Richard}, {Maseda}, {Bouwens}, {Bouch{\'e}}, {Kollatschny}, {Schaye}, {Marino}, {Pello}, {Herenz}, {Guiderdoni}, \& {Carollo}}]{MUSE_DR1_2017}
{Bacon}, R., {Conseil}, S., {Mary}, D., {et~al.} 2017, \aap, 608, A1

\bibitem[{{Barrera-Ballesteros} {et~al.}(2014){Barrera-Ballesteros}, {Falc{\'o}n-Barroso}, {Garc{\'\i}a-Lorenzo}, {van de Ven}, {Aguerri}, {Mendez-Abreu}, {Spekkens}, {Lyubenova}, {S{\'a}nchez}, {Husemann}, {Mast}, {Garc{\'\i}a-Benito}, {Iglesias-Paramo}, {Del Olmo}, {M{\'a}rquez}, {Masegosa}, {Kehrig}, {Marino}, {Verdes-Montenegro}, {Ziegler}, {McIntosh}, {Bland-Hawthorn}, {Walcher}, \& {CALIFA Collaboration}}]{Barrera-Ballesteros2014}
{Barrera-Ballesteros}, J.~K., {Falc{\'o}n-Barroso}, J., {Garc{\'\i}a-Lorenzo}, B., {et~al.} 2014, \aap, 568, A70

\bibitem[{{Beckwith} {et~al.}(2006){Beckwith}, {Stiavelli}, {Koekemoer}, {Caldwell}, {Ferguson}, {Hook}, {Lucas}, {Bergeron}, {Corbin}, {Jogee}, {Panagia}, {Robberto}, {Royle}, {Somerville}, \& {Sosey}}]{HUDF_2006}
{Beckwith}, S. V.~W., {Stiavelli}, M., {Koekemoer}, A.~M., {et~al.} 2006, \aj, 132, 1729

\bibitem[{{Bekki}(2015)}]{Bekki_2015}
{Bekki}, K. 2015, \mnras, 454, L41

\bibitem[{{Bergvall} {et~al.}(2006){Bergvall}, {Zackrisson}, {Andersson}, {Arnberg}, {Masegosa}, \& {{\"O}stlin}}]{Bergvall_2006}
{Bergvall}, N., {Zackrisson}, E., {Andersson}, B.~G., {et~al.} 2006, \aap, 448, 513

\bibitem[{{Bittner} {et~al.}(2019){Bittner}, {Falc{\'o}n-Barroso}, {Nedelchev}, {Dorta}, {Gadotti}, {Sarzi}, {Molaeinezhad}, {Iodice}, {Rosado-Belza}, {de Lorenzo-C{\'a}ceres}, {Fragkoudi}, {Gal{\'a}n-de Anta}, {Husemann}, {M{\'e}ndez-Abreu}, {Neumann}, {Pinna}, {Querejeta}, {S{\'a}nchez-Bl{\'a}zquez}, \& {Seidel}}]{Bittner_2019}
{Bittner}, A., {Falc{\'o}n-Barroso}, J., {Nedelchev}, B., {et~al.} 2019, \aap, 628, A117

\bibitem[{{Bosch} {et~al.}(2019){Bosch}, {H{\"a}gele}, {Amor{\'\i}n}, {Firpo}, {Cardaci}, {V{\'\i}lchez}, {P{\'e}rez-Montero}, {Papaderos}, {Dors}, {Krabbe}, \& {Campuzano-Castro}}]{Bosch_2019}
{Bosch}, G., {H{\"a}gele}, G.~F., {Amor{\'\i}n}, R., {et~al.} 2019, \mnras, 489, 1787

\bibitem[{{Brinchmann} {et~al.}(2004){Brinchmann}, {Charlot}, {White}, {Tremonti}, {Kauffmann}, {Heckman}, \& {Brinkmann}}]{Brinchmann_2004}
{Brinchmann}, J., {Charlot}, S., {White}, S.~D.~M., {et~al.} 2004, \mnras, 351, 1151

\bibitem[{{Bruzual} \& {Charlot}(2003)}]{Bruzual_Charlot_2003}
{Bruzual}, G. \& {Charlot}, S. 2003, \mnras, 344, 1000

\bibitem[{{Cardamone} {et~al.}(2009){Cardamone}, {Schawinski}, {Sarzi}, {Bamford}, {Bennert}, {Urry}, {Lintott}, {Keel}, {Parejko}, {Nichol}, {Thomas}, {Andreescu}, {Murray}, {Raddick}, {Slosar}, {Szalay}, \& {Vandenberg}}]{Cardamone_2009}
{Cardamone}, C., {Schawinski}, K., {Sarzi}, M., {et~al.} 2009, \mnras, 399, 1191

\bibitem[{{Cardelli} {et~al.}(1989){Cardelli}, {Clayton}, \& {Mathis}}]{Cardelli_1989}
{Cardelli}, J.~A., {Clayton}, G.~C., \& {Mathis}, J.~S. 1989, \apj, 345, 245

\bibitem[{{Dayal} \& {Ferrara}(2018)}]{Dayal_2018}
{Dayal}, P. \& {Ferrara}, A. 2018, \physrep, 780, 1

\bibitem[{{del Moral-Castro} {et~al.}(2019){del Moral-Castro}, {Garc{\'\i}a-Lorenzo}, {Ramos Almeida}, {Ruiz-Lara}, {Falc{\'o}n-Barroso}, {S{\'a}nchez}, {S{\'a}nchez-Bl{\'a}zquez}, {M{\'a}rquez}, \& {Masegosa}}]{delMoral2019}
{del Moral-Castro}, I., {Garc{\'\i}a-Lorenzo}, B., {Ramos Almeida}, C., {et~al.} 2019, \mnras, 485, 3794

\bibitem[{{del Moral-Castro} {et~al.}(2020){del Moral-Castro}, {Garc{\'\i}a-Lorenzo}, {Ramos Almeida}, {Ruiz-Lara}, {Falc{\'o}n-Barroso}, {S{\'a}nchez}, {S{\'a}nchez-Bl{\'a}zquez}, {M{\'a}rquez}, \& {Masegosa}}]{delMoral2020}
{del Moral-Castro}, I., {Garc{\'\i}a-Lorenzo}, B., {Ramos Almeida}, C., {et~al.} 2020, \aap, 639, L9

\bibitem[{{Dom{\'\i}nguez-G{\'o}mez} {et~al.}(2023){Dom{\'\i}nguez-G{\'o}mez}, {P{\'e}rez}, {Ruiz-Lara}, {Peletier}, {S{\'a}nchez-Bl{\'a}zquez}, {Lisenfeld}, {Falc{\'o}n-Barroso}, {Alc{\'a}zar-Laynez}, {Argudo-Fern{\'a}ndez}, {Bl{\'a}zquez-Calero}, {Courtois}, {Duarte Puertas}, {Espada}, {Florido}, {Garc{\'\i}a-Benito}, {Jim{\'e}nez}, {Kreckel}, {Rela{\~n}o}, {S{\'a}nchez-Menguiano}, {van der Hulst}, {van de Weygaert}, {Verley}, \& {Zurita}}]{Jesus_2023Natur}
{Dom{\'\i}nguez-G{\'o}mez}, J., {P{\'e}rez}, I., {Ruiz-Lara}, T., {et~al.} 2023, \nat, 619, 269

\bibitem[{{Duarte Puertas} {et~al.}(2022){Duarte Puertas}, {Vilchez}, {Iglesias-P{\'a}ramo}, {Moll{\'a}}, {P{\'e}rez-Montero}, {Kehrig}, {Pilyugin}, \& {Zinchenko}}]{Duarte_2022}
{Duarte Puertas}, S., {Vilchez}, J.~M., {Iglesias-P{\'a}ramo}, J., {et~al.} 2022, \aap, 666, A186

\bibitem[{{Erb} {et~al.}(2016){Erb}, {Pettini}, {Steidel}, {Strom}, {Rudie}, {Trainor}, {Shapley}, \& {Reddy}}]{Erb_2016}
{Erb}, D.~K., {Pettini}, M., {Steidel}, C.~C., {et~al.} 2016, \apj, 830, 52

\bibitem[{{Fletcher} {et~al.}(2019){Fletcher}, {Tang}, {Robertson}, {Nakajima}, {Ellis}, {Stark}, \& {Inoue}}]{Fletcher_2019}
{Fletcher}, T.~J., {Tang}, M., {Robertson}, B.~E., {et~al.} 2019, \apj, 878, 87

\bibitem[{{Gu{\'e}rou} {et~al.}(2017){Gu{\'e}rou}, {Krajnovi{\'c}}, {Epinat}, {Contini}, {Emsellem}, {Bouch{\'e}}, {Bacon}, {Michel-Dansac}, {Richard}, {Weilbacher}, {Schaye}, {Marino}, {den Brok}, \& {Erroz-Ferrer}}]{Guerou_2017}
{Gu{\'e}rou}, A., {Krajnovi{\'c}}, D., {Epinat}, B., {et~al.} 2017, \aap, 608, A5

\bibitem[{{Haro}(1956)}]{Haro_1956}
{Haro}, G. 1956, Boletin de los Observatorios Tonantzintla y Tacubaya, 2, 8

\bibitem[{{Harshan} {et~al.}(2020){Harshan}, {Gupta}, {Tran}, {Alcorn}, {Yuan}, {Kacprzak}, {Nanayakkara}, {Glazebrook}, {Kewley}, {Labb{\'e}}, \& {Papovich}}]{Harshan_2020}
{Harshan}, A., {Gupta}, A., {Tran}, K.-V., {et~al.} 2020, \apj, 892, 77

\bibitem[{Hunter(2007)}]{Hunter_2007}
Hunter, J.~D. 2007, Computing in Science \& Engineering, 9, 90

\bibitem[{{Iglesias-P{\'a}ramo} {et~al.}(2022){Iglesias-P{\'a}ramo}, {Arroyo}, {Kehrig}, {V{\'\i}lchez}, {Duarte Puertas}, {P{\'e}rez-Montero}, {Breda}, {Jim{\'e}nez-Teja}, {L{\'o}pez Sanjuan}, {Lumbreras-Calle}, {Coelho}, {Gurung-L{\'o}pez}, {Queiroz}, {M{\'a}rquez}, {Povi{\'c}}, {Gonz{\'a}lez Delgado}, {Chaves-Montero}, {Sobral}, {Hern{\'a}n-Caballero}, {Fern{\'a}ndez-Ontiveros}, {D{\'\i}az-Garc{\'\i}a}, {Alvarez-Candal}, {Abramo}, {Alcaniz}, {Ben{\'\i}tez}, {Bonoli}, {Cenarro}, {Crist{\'o}bal-Hornillos}, {Dupke}, {Ederoclite}, {Mar{\'\i}n-Franch}, {Mendes de Oliveira}, {Moles}, {Sodr{\'e}}, {Taylor}, {Varela}, {V{\'a}zquez Rami{\'o}}, \& {J-PAS Team}}]{Jorge_2022_EELG_JPAS}
{Iglesias-P{\'a}ramo}, J., {Arroyo}, A., {Kehrig}, C., {et~al.} 2022, \aap, 665, A95

\bibitem[{{Inami} {et~al.}(2017){Inami}, {Bacon}, {Brinchmann}, {Richard}, {Contini}, {Conseil}, {Hamer}, {Akhlaghi}, {Bouch{\'e}}, {Cl{\'e}ment}, {Desprez}, {Drake}, {Hashimoto}, {Leclercq}, {Maseda}, {Michel-Dansac}, {Paalvast}, {Tresse}, {Ventou}, {Kollatschny}, {Boogaard}, {Finley}, {Marino}, {Schaye}, \& {Wisotzki}}]{Inami_DR1_2017}
{Inami}, H., {Bacon}, R., {Brinchmann}, J., {et~al.} 2017, \aap, 608, A2

\bibitem[{{Indahl} {et~al.}(2021){Indahl}, {Zeimann}, {Hill}, {Bowman}, {Ciardullo}, {Drory}, {Gawiser}, {Hopp}, {Janowiecki}, {Boylan-Kolchin}, {Mentuch Cooper}, {Davis}, {Farrow}, {Finkelstein}, {Gronwall}, {Kelz}, {McQuinn}, {Schneider}, \& {Tuttle}}]{Indahl_2021}
{Indahl}, B., {Zeimann}, G., {Hill}, G.~J., {et~al.} 2021, \apj, 916, 11

\bibitem[{{Izotov} {et~al.}(2019){Izotov}, {Guseva}, {Fricke}, \& {Henkel}}]{Izotov_2019}
{Izotov}, Y.~I., {Guseva}, N.~G., {Fricke}, K.~J., \& {Henkel}, C. 2019, \aap, 623, A40

\bibitem[{{Izotov} {et~al.}(2016){Izotov}, {Schaerer}, {Thuan}, {Worseck}, {Guseva}, {Orlitov{\'a}}, \& {Verhamme}}]{Izotov_2016}
{Izotov}, Y.~I., {Schaerer}, D., {Thuan}, T.~X., {et~al.} 2016, \mnras, 461, 3683

\bibitem[{{Izotov} {et~al.}(2018){Izotov}, {Worseck}, {Schaerer}, {Guseva}, {Thuan}, {Fricke}, \& {Orlitov{\'a}}}]{Izotov_2018}
{Izotov}, Y.~I., {Worseck}, G., {Schaerer}, D., {et~al.} 2018, \mnras, 478, 4851

\bibitem[{{Jaskot} \& {Oey}(2013)}]{JaskotOey_2013}
{Jaskot}, A.~E. \& {Oey}, M.~S. 2013, \apj, 766, 91

\bibitem[{{Kehrig} {et~al.}(2016){Kehrig}, {V{\'\i}lchez}, {P{\'e}rez-Montero}, {Iglesias-P{\'a}ramo}, {Hern{\'a}ndez-Fern{\'a}ndez}, {Duarte Puertas}, {Brinchmann}, {Durret}, \& {Kunth}}]{Kehrig_2016}
{Kehrig}, C., {V{\'\i}lchez}, J.~M., {P{\'e}rez-Montero}, E., {et~al.} 2016, \mnras, 459, 2992

\bibitem[{{Lagos} {et~al.}(2016){Lagos}, {Demarco}, {Papaderos}, {Telles}, {Nigoche-Netro}, {Humphrey}, {Roche}, \& {Gomes}}]{Lagos_2016}
{Lagos}, P., {Demarco}, R., {Papaderos}, P., {et~al.} 2016, \mnras, 456, 1549

\bibitem[{{Lofthouse} {et~al.}(2017){Lofthouse}, {Houghton}, \& {Kaviraj}}]{Lofthouse_2017}
{Lofthouse}, E.~K., {Houghton}, R.~C.~W., \& {Kaviraj}, S. 2017, \mnras, 471, 2311

\bibitem[{{Lumbreras-Calle} {et~al.}(2022){Lumbreras-Calle}, {L{\'o}pez-Sanjuan}, {Sobral}, {Fern{\'a}ndez-Ontiveros}, {V{\'\i}lchez}, {Hern{\'a}n-Caballero}, {Akhlaghi}, {D{\'\i}az-Garc{\'\i}a}, {Alcaniz}, {Angulo}, {Cenarro}, {Crist{\'o}bal-Hornillos}, {Dupke}, {Ederoclite}, {Hern{\'a}ndez-Monteagudo}, {Mar{\'\i}n-Franch}, {Moles}, {Sodr{\'e}}, {V{\'a}zquez Rami{\'o}}, \& {Varela}}]{Lumbreras-Calle_2022}
{Lumbreras-Calle}, A., {L{\'o}pez-Sanjuan}, C., {Sobral}, D., {et~al.} 2022, \aap, 668, A60

\bibitem[{{Markarian}(1967)}]{Markarian_1967}
{Markarian}, B.~E. 1967, Astrofizika, 3, 24

\bibitem[{{Matthee} {et~al.}(2022){Matthee}, {Naidu}, {Pezzulli}, {Gronke}, {Sobral}, {Oesch}, {Hayes}, {Erb}, {Schaerer}, {Amor{\'\i}n}, {Tacchella}, {Paulino-Afonso}, {Llerena}, {Calhau}, \& {R{\"o}ttgering}}]{Matthee_2022}
{Matthee}, J., {Naidu}, R.~P., {Pezzulli}, G., {et~al.} 2022, \mnras, 512, 5960

\bibitem[{{Naidu} {et~al.}(2022){Naidu}, {Matthee}, {Oesch}, {Conroy}, {Sobral}, {Pezzulli}, {Hayes}, {Erb}, {Amor{\'\i}n}, {Gronke}, {Schaerer}, {Tacchella}, {Kerutt}, {Paulino-Afonso}, {Calhau}, {Llerena}, \& {R{\"o}ttgering}}]{Naidu_2022}
{Naidu}, R.~P., {Matthee}, J., {Oesch}, P.~A., {et~al.} 2022, \mnras, 510, 4582

\bibitem[{{Paalvast} {et~al.}(2018){Paalvast}, {Verhamme}, {Straka}, {Brinchmann}, {Herenz}, {Carton}, {Gunawardhana}, {Boogaard}, {Cantalupo}, {Contini}, {Epinat}, {Inami}, {Marino}, {Maseda}, {Michel-Dansac}, {Muzahid}, {Nanayakkara}, {Pezzulli}, {Richard}, {Schaye}, {Segers}, {Urrutia}, {Wendt}, \& {Wisotzki}}]{Paalvast_2018}
{Paalvast}, M., {Verhamme}, A., {Straka}, L.~A., {et~al.} 2018, \aap, 618, A40

\bibitem[{{Pagel} {et~al.}(1979){Pagel}, {Edmunds}, {Blackwell}, {Chun}, \& {Smith}}]{Pagel_1979}
{Pagel}, B.~E.~J., {Edmunds}, M.~G., {Blackwell}, D.~E., {Chun}, M.~S., \& {Smith}, G. 1979, \mnras, 189, 95

\bibitem[{{P{\'e}rez-Montero}(2017)}]{EPM_2017PASP}
{P{\'e}rez-Montero}, E. 2017, \pasp, 129, 043001

\bibitem[{{P{\'e}rez-Montero} {et~al.}(2021){P{\'e}rez-Montero}, {Amor{\'\i}n}, {S{\'a}nchez Almeida}, {V{\'\i}lchez}, {Garc{\'\i}a-Benito}, \& {Kehrig}}]{Perez_montero_2021}
{P{\'e}rez-Montero}, E., {Amor{\'\i}n}, R., {S{\'a}nchez Almeida}, J., {et~al.} 2021, \mnras, 504, 1237

\bibitem[{{P{\'e}rez-Montero} {et~al.}(2020){P{\'e}rez-Montero}, {Kehrig}, {V{\'\i}lchez}, {Garc{\'\i}a-Benito}, {Duarte Puertas}, \& {Iglesias-P{\'a}ramo}}]{Perez_montero_2020}
{P{\'e}rez-Montero}, E., {Kehrig}, C., {V{\'\i}lchez}, J.~M., {et~al.} 2020, \aap, 643, A80

\bibitem[{{Ravindranath} {et~al.}(2020){Ravindranath}, {Monroe}, {Jaskot}, {Ferguson}, \& {Tumlinson}}]{Ravindranath_2020}
{Ravindranath}, S., {Monroe}, T., {Jaskot}, A., {Ferguson}, H.~C., \& {Tumlinson}, J. 2020, \apj, 896, 170

\bibitem[{{Rela{\~n}o} {et~al.}(2010){Rela{\~n}o}, {Monreal-Ibero}, {V{\'\i}lchez}, \& {Kennicutt}}]{Relano_2010}
{Rela{\~n}o}, M., {Monreal-Ibero}, A., {V{\'\i}lchez}, J.~M., \& {Kennicutt}, R.~C. 2010, \mnras, 402, 1635

\bibitem[{{Roche} {et~al.}(2023){Roche}, {V{\'\i}lchez}, {Iglesias-P{\'a}ramo}, {Papaderos}, {S{\'a}nchez}, {Kehrig}, \& {Duarte Puertas}}]{Roche_2023}
{Roche}, N., {V{\'\i}lchez}, J.~M., {Iglesias-P{\'a}ramo}, J., {et~al.} 2023, \mnras, 523, 270

\bibitem[{{Salzer} {et~al.}(2020){Salzer}, {Feddersen}, {Derloshon}, {Gronwall}, {Van Sistine}, {Sugden}, {Janowiecki}, {Hirschauer}, \& {Kellar}}]{Salzer_2020}
{Salzer}, J.~J., {Feddersen}, J.~R., {Derloshon}, K., {et~al.} 2020, \aj, 160, 242

\bibitem[{{Sanders} {et~al.}(2016){Sanders}, {Shapley}, {Kriek}, {Reddy}, {Freeman}, {Coil}, {Siana}, {Mobasher}, {Shivaei}, {Price}, \& {de Groot}}]{Sanders_2016}
{Sanders}, R.~L., {Shapley}, A.~E., {Kriek}, M., {et~al.} 2016, \apj, 816, 23

\bibitem[{{Schaerer} {et~al.}(2022){Schaerer}, {Marques-Chaves}, {Barrufet}, {Oesch}, {Izotov}, {Naidu}, {Guseva}, \& {Brammer}}]{Schaerer_2022}
{Schaerer}, D., {Marques-Chaves}, R., {Barrufet}, L., {et~al.} 2022, \aap, 665, L4

\bibitem[{{Schlafly} \& {Finkbeiner}(2011)}]{Extinct_NED2011}
{Schlafly}, E.~F. \& {Finkbeiner}, D.~P. 2011, \apj, 737, 103

\bibitem[{{Stark}(2016)}]{Stark_2016}
{Stark}, D.~P. 2016, \araa, 54, 761

\bibitem[{{Storey} \& {Hummer}(1995)}]{StoreyHummer1995}
{Storey}, P.~J. \& {Hummer}, D.~G. 1995, \mnras, 272, 41

\bibitem[{{Sun} {et~al.}(2023){Sun}, {Egami}, {Pirzkal}, {Rieke}, {Baum}, {Boyer}, {Boyett}, {Bunker}, {Cameron}, {Curti}, {Eisenstein}, {Gennaro}, {Greene}, {Jaffe}, {Kelly}, {Koekemoer}, {Kumari}, {Maiolino}, {Maseda}, {Perna}, {Rest}, {Robertson}, {Schlawin}, {Smit}, {Stansberry}, {Sunnquist}, {Tacchella}, {Williams}, \& {Willmer}}]{Sun_2023}
{Sun}, F., {Egami}, E., {Pirzkal}, N., {et~al.} 2023, \apj, 953, 53

\bibitem[{{Tang} {et~al.}(2021){Tang}, {Stark}, {Chevallard}, {Charlot}, {Endsley}, \& {Congiu}}]{Tang_2021}
{Tang}, M., {Stark}, D.~P., {Chevallard}, J., {et~al.} 2021, \mnras, 503, 4105

\bibitem[{{Terlevich} {et~al.}(1991){Terlevich}, {Melnick}, {Masegosa}, {Moles}, \& {Copetti}}]{Terlevich_1991}
{Terlevich}, R., {Melnick}, J., {Masegosa}, J., {Moles}, M., \& {Copetti}, M.~V.~F. 1991, \aaps, 91, 285

\bibitem[{{Williams} {et~al.}(2023){Williams}, {Kelly}, {Chen}, {Brammer}, {Zitrin}, {Treu}, {Scarlata}, {Koekemoer}, {Oguri}, {Lin}, {Diego}, {Nonino}, {Hjorth}, {Langeroodi}, {Broadhurst}, {Rogers}, {Perez-Fournon}, {Foley}, {Jha}, {Filippenko}, {Strolger}, {Pierel}, {Poidevin}, \& {Yang}}]{Williams_2023}
{Williams}, H., {Kelly}, P.~L., {Chen}, W., {et~al.} 2023, Science, 380, 416

\bibitem[{{Yang} {et~al.}(2017{\natexlab{a}}){Yang}, {Malhotra}, {Gronke}, {Rhoads}, {Leitherer}, {Wofford}, {Jiang}, {Dijkstra}, {Tilvi}, \& {Wang}}]{2017ApJ...844..171Y}
{Yang}, H., {Malhotra}, S., {Gronke}, M., {et~al.} 2017{\natexlab{a}}, \apj, 844, 171

\bibitem[{{Yang} {et~al.}(2017{\natexlab{b}}){Yang}, {Malhotra}, {Rhoads}, \& {Wang}}]{Yang_2017}
{Yang}, H., {Malhotra}, S., {Rhoads}, J.~E., \& {Wang}, J. 2017{\natexlab{b}}, \apj, 847, 38

\bibitem[{{Zwicky}(1966)}]{Zwicky_1966}
{Zwicky}, F. 1966, \apj, 143, 192

\end{thebibliography}
%

\begin{appendix}

\section{Spectra and images of the primary  EELG sample candidates}\label{sec:App_EELGs_HST_images}

In this appendix, we show both the spectra and the images of the selected EELG candidates.


\begin{figure*}
\begin{center}

\includegraphics[width=\textwidth]{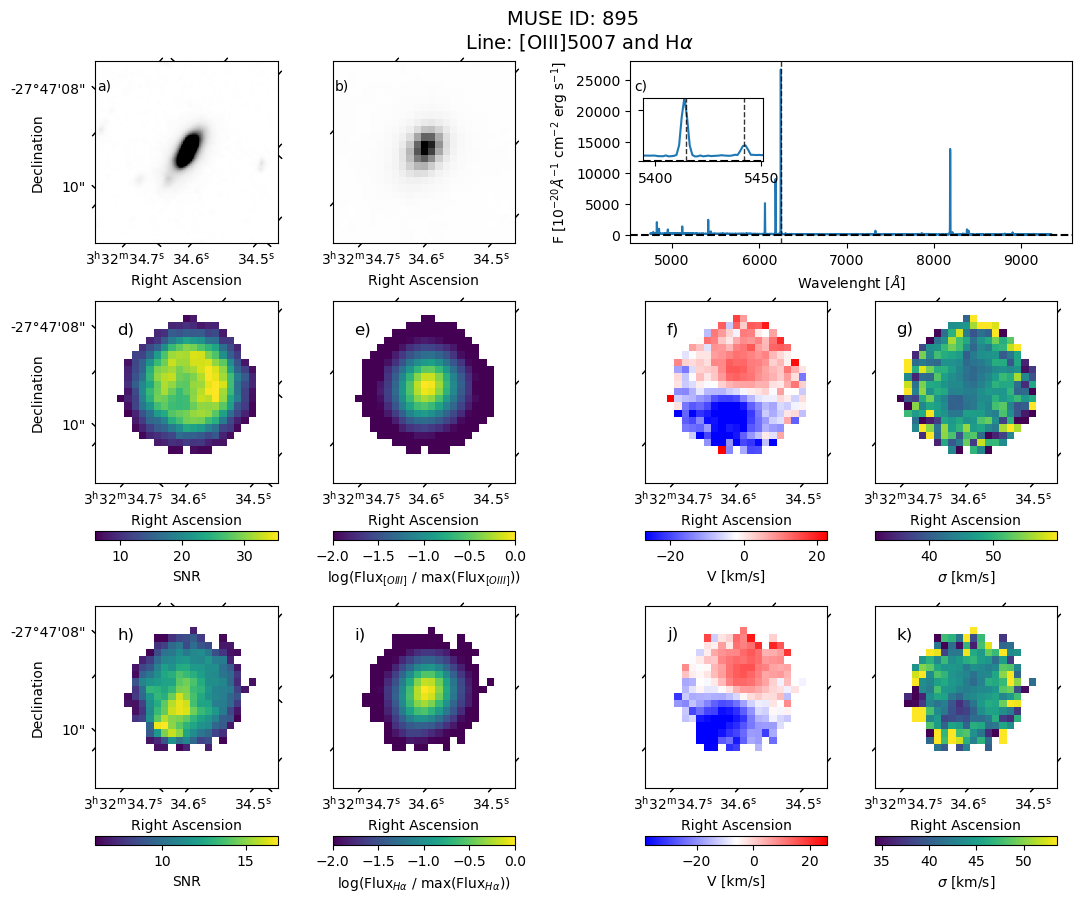}

\caption[]{Same as Fig. \ref{fig:HST_white-light_spectrum_maps_891} for the MUSE ID galaxy 895.} \label{fig:SNR_flux_vel_sig_maps_895} 
 \end{center}
\end{figure*}

\begin{figure*}
\begin{center}

\includegraphics[width=\textwidth]{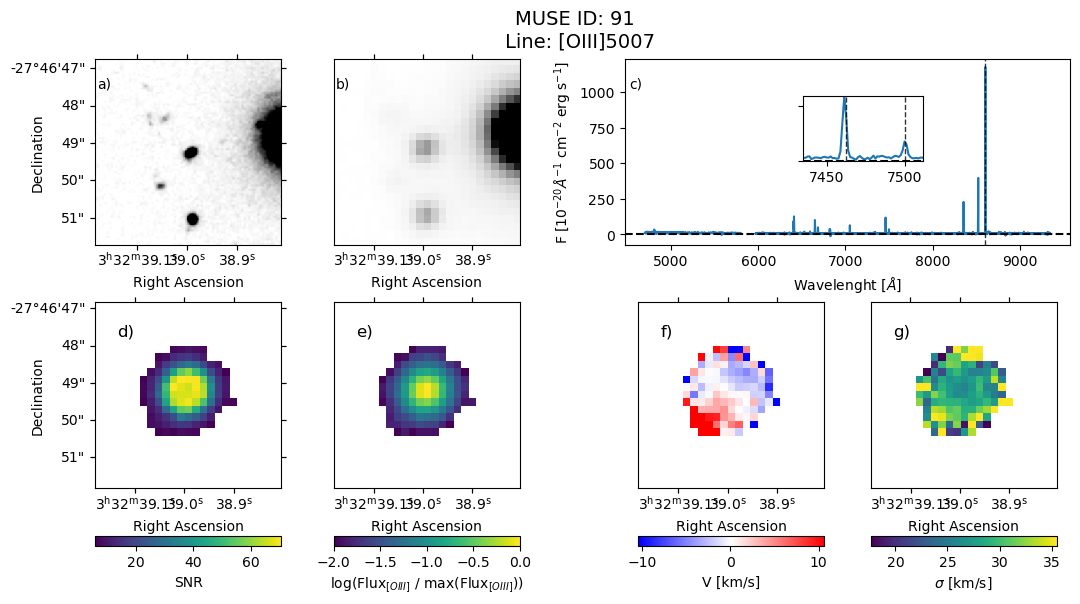}

\caption[]{Same as Fig. \ref{fig:HST_white-light_spectrum_maps_891} for the MUSE ID galaxy 91.} \label{fig:SNR_flux_vel_sig_maps_91} 
 \end{center}
\end{figure*}

\begin{figure*}
\begin{center}

\includegraphics[width=\textwidth]{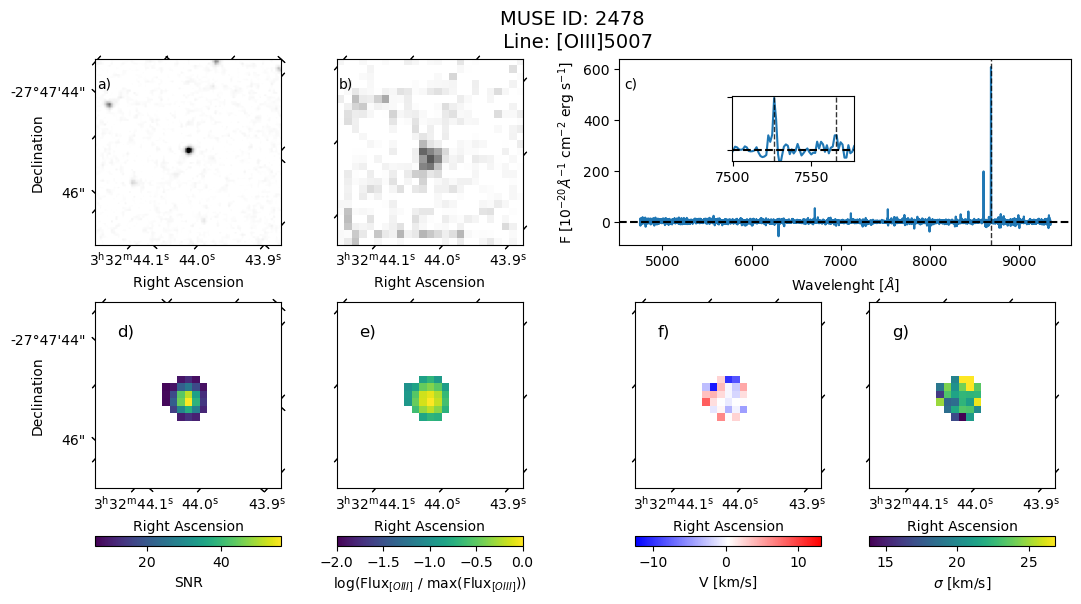}

\caption[]{Same as Fig. \ref{fig:HST_white-light_spectrum_maps_891} for the MUSE ID galaxy 2478.} \label{fig:SNR_flux_vel_sig_maps_2478} 
 \end{center}
\end{figure*}

\begin{figure*}
\begin{center}

\includegraphics[width=\textwidth]{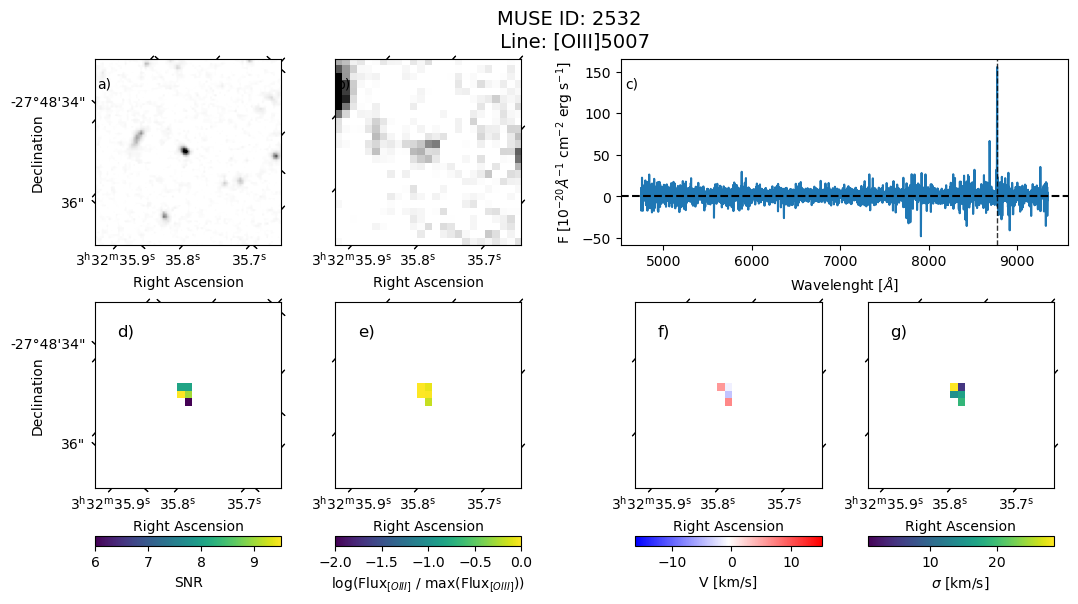}

\caption[]{Same as Fig. \ref{fig:HST_white-light_spectrum_maps_891} for the MUSE ID galaxy 2532.} \label{fig:SNR_flux_vel_sig_maps_2532} 
 \end{center}
\end{figure*}

\begin{figure*}
\begin{center}

\includegraphics[width=\textwidth]{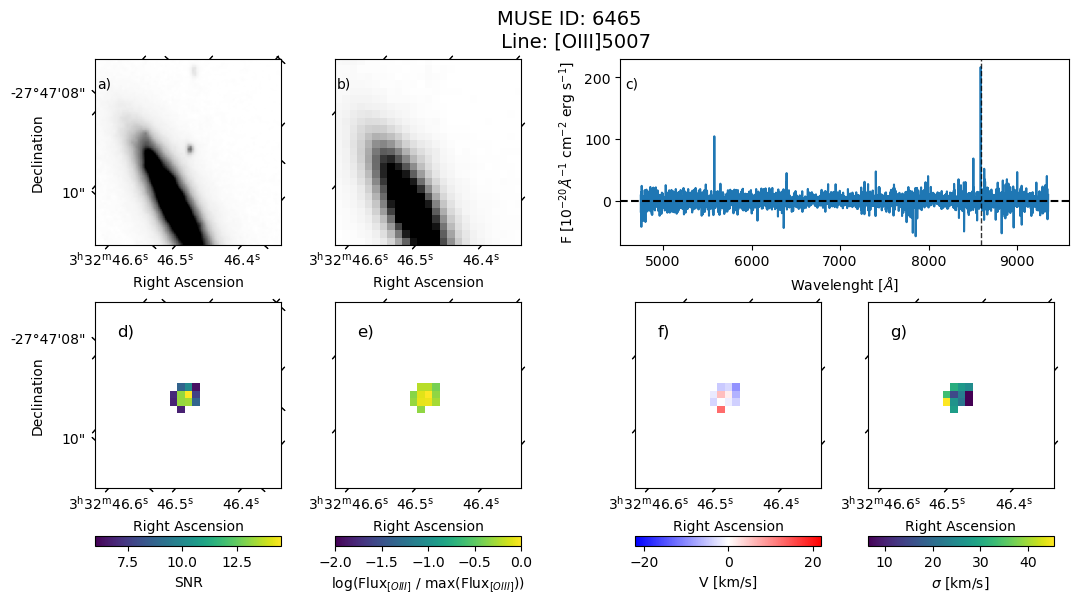}

\caption[]{Same as Fig. \ref{fig:HST_white-light_spectrum_maps_891} for the MUSE ID galaxy 6465.} \label{fig:SNR_flux_vel_sig_maps_6465} 
 \end{center}
\end{figure*}

\begin{figure*}
\begin{center}

\includegraphics[width=\textwidth]{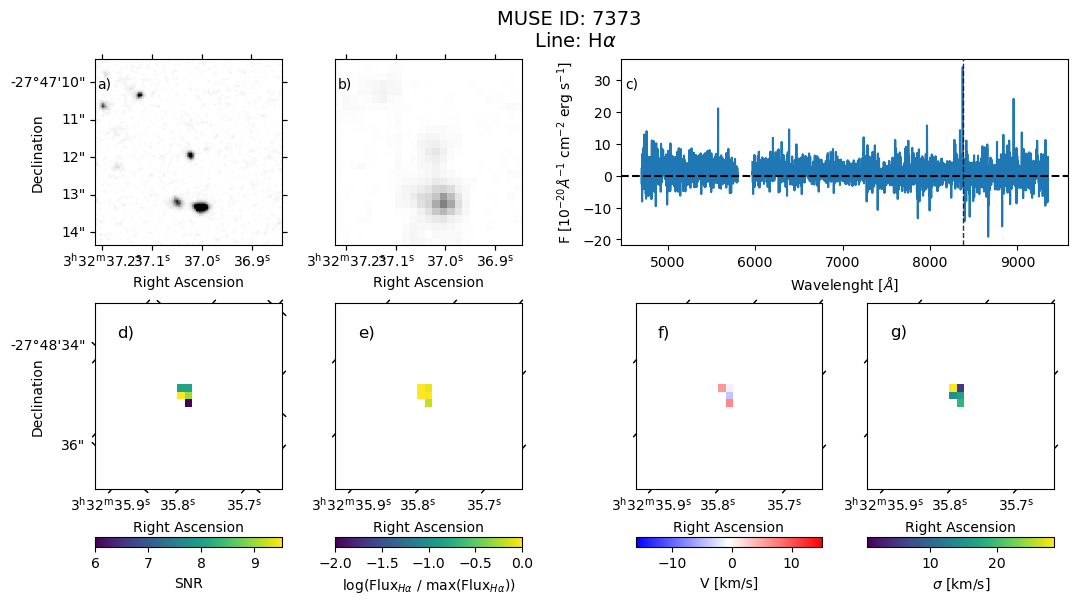}

\caption[]{Same as Fig. \ref{fig:HST_white-light_spectrum_maps_891} for the MUSE ID galaxy 7373.} \label{fig:flux_vel_maps_Ha_OIII_7373} 
 \end{center}
\end{figure*}


\section{Spectra and images of the EELG extended sample}\label{sec:App_extended_HST_images}

Here, we present both the spectra and the images of the extended sample.

\begin{figure*}
\begin{center}

\includegraphics[width=\textwidth]{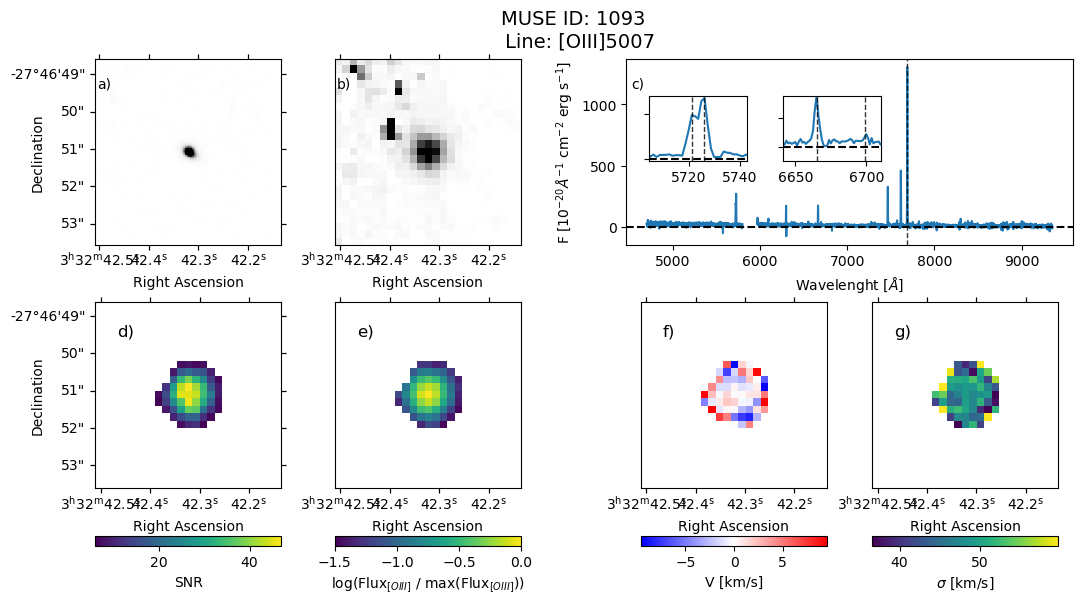}

\caption[]{Same as Fig. \ref{fig:HST_white-light_spectrum_maps_891} for the MUSE ID galaxy 1093.} \label{fig:flux_vel_maps_Ha_OIII_1093} 
 \end{center}
\end{figure*}

\begin{figure*}
\begin{center} 

\includegraphics[width=\textwidth]{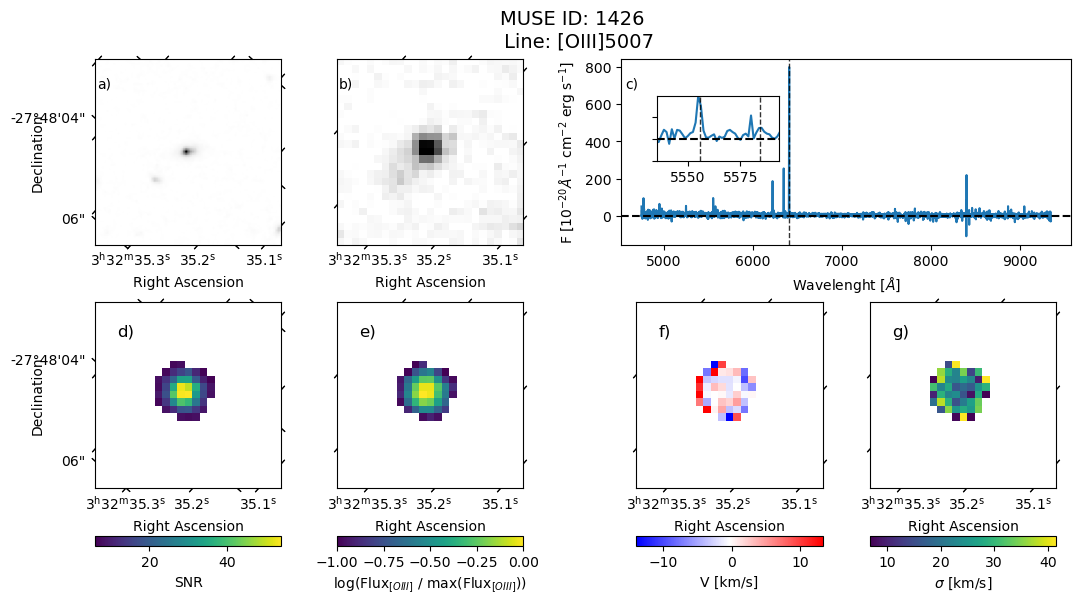}

\caption[]{Same as Fig. \ref{fig:HST_white-light_spectrum_maps_891} for the MUSE ID galaxy 1426.} \label{fig:flux_vel_maps_Ha_OIII_1426} 
 \end{center}
\end{figure*}

\begin{figure*}
\begin{center}

\includegraphics[width=\textwidth]{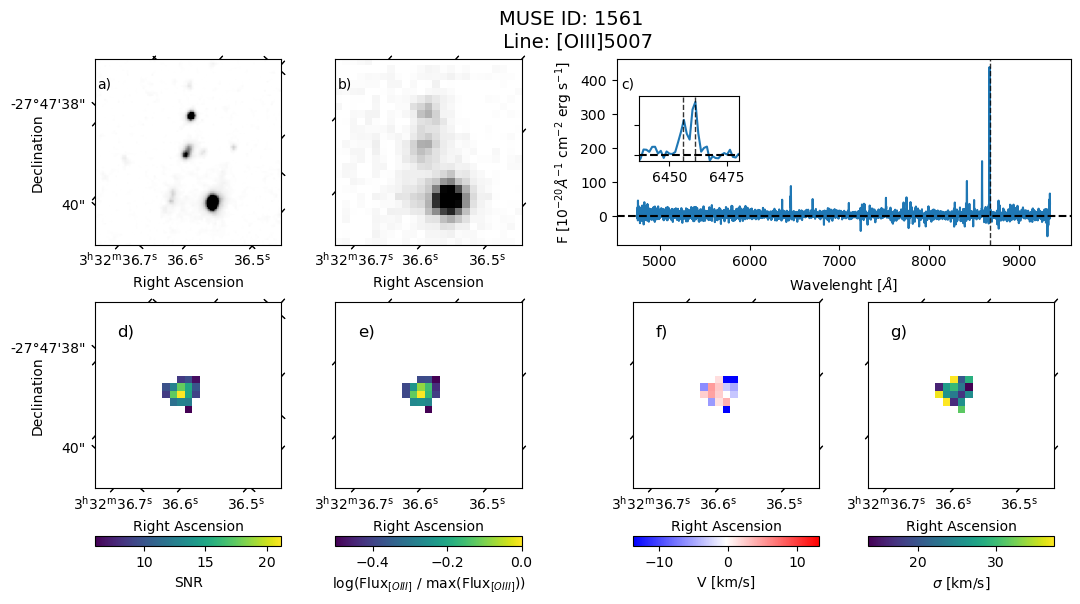}

\caption[]{Same as Fig. \ref{fig:HST_white-light_spectrum_maps_891} for the MUSE ID galaxy 1561.} \label{fig:flux_vel_maps_Ha_OIII_1561} 
 \end{center}
\end{figure*}

\begin{figure*}
\begin{center}

\includegraphics[width=\textwidth]{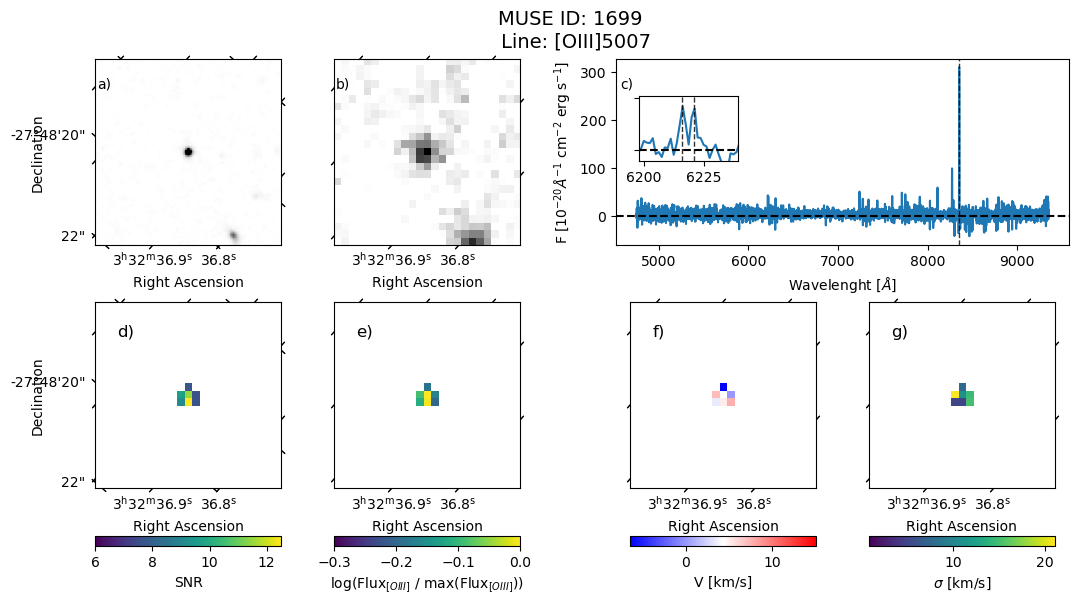}

\caption[]{Same as Fig. \ref{fig:HST_white-light_spectrum_maps_891} for the MUSE ID galaxy 1699.} \label{fig:flux_vel_maps_Ha_OIII_1699} 
 \end{center}
\end{figure*}

\begin{figure*}
\begin{center}

\includegraphics[width=\textwidth]{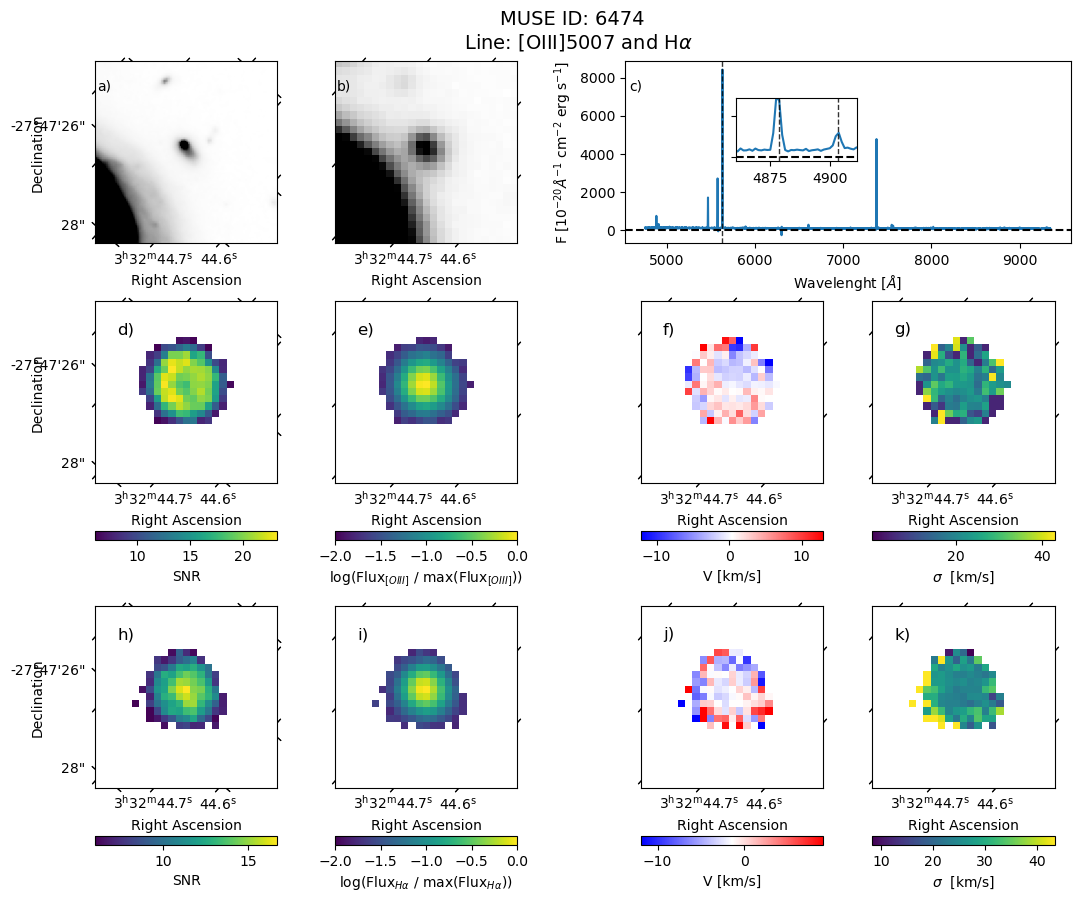}

\caption[]{Same as Fig. \ref{fig:HST_white-light_spectrum_maps_891} for the MUSE ID galaxy 6474.} \label{fig:flux_vel_maps_Ha_OIII_6474} 
 \end{center}
\end{figure*}

\begin{figure*}
\begin{center}

\includegraphics[width=\textwidth]{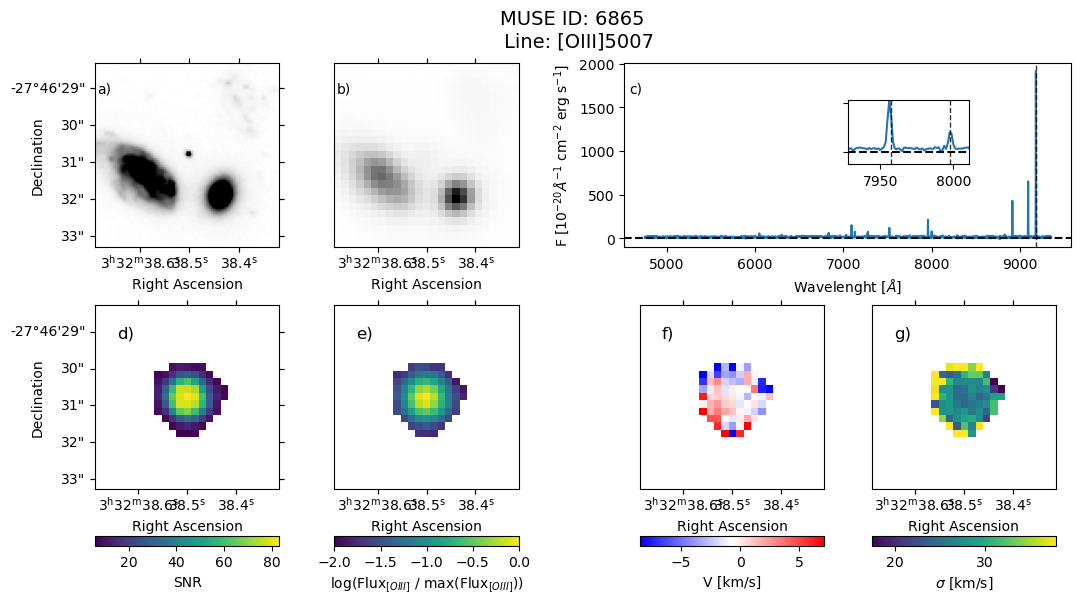}

\caption[]{Same as Fig. \ref{fig:HST_white-light_spectrum_maps_891} for the MUSE ID galaxy 6865.} \label{fig:flux_vel_maps_Ha_OIII_6865} 
 \end{center}
\end{figure*}

\section{Line fluxes of the EELG primary sample candidates}\label{sec:App_fluxes_EELG}

In this appendix, we show the fluxes and equivalent width measured with pyPlatefit for our EELG sample. Only lines with SNR higher than 3 are included in the tables.

\begin{table*}
\begin{center}
\begin{tabular}{ c c | c|  c | c|  c|  c|  c | c }
\hline
\hline
      & Galaxy ID          & 91    & 891   & 895  & 2478  & 2532  & 6465 & 7373 \\
\hline
\hline
Line  & $\lambda$ & F & F & F & F & F & F & F \\
      &    (\AA{})      &  (10$^{-20}\frac{erg}{s cm^{2}}$) & (10$^{-20}\frac{erg}{s cm^{2}}$)  &  (10$^{-20}\frac{erg}{s cm^{2}}$) & (10$^{-20}\frac{erg}{s cm^{2}}$)  &  (10$^{-20}\frac{erg}{s cm^{2}}$) & (10$^{-20}\frac{erg}{s cm^{2}}$) &  (10$^{-20}\frac{erg}{s cm^{2}}$) \\
   (1)      &    (2)      & (3)  & (4)  & (5)  &  (6)    &  (7) & (8) & (9)  \\
\hline
$[$OII$]$ & 3726.03 & 257 $\pm$ 7 & ... & ... & ... & ... & 94 $\pm$ 2 & ... \\
$[$OII$]$ & 3728.82 & 356 $\pm$ 5 & ... & ... & ... & ... & 125 $\pm$ 3 & ... \\
H11 & 3770.63 & 21 $\pm$ 5 & ... & ... & ... & ... & ... & ... \\
H10 & 3797.90 & 41 $\pm$ 4 & ... & ... & ... & ... & ... & ... \\
H9 & 3835.39 & 52 $\pm$ 3 & ... & 1347 $\pm$ 34 & ... & ... & ... & ... \\
$[$NeIII$]$ & 3868.75 & 277 $\pm$ 8 & ... & 6342 $\pm$ 8 & 135 $\pm$ 4 & 38 $\pm$ 4 & 54 $\pm$ 2 & ... \\
$[$HeI$]$ & 3888.65 & 132 $\pm$ 3 & 4266 $\pm$ 8 & 3298 $\pm$ 28 & 48 $\pm$ 2 & ... & ... & ... \\
H8 & 3889.05 & 134 $\pm$ 5 & 4213 $\pm$ 8 & 3264 $\pm$ 28 & 47 $\pm$ 2 & ... & ... & ... \\
$[$NeIII$]$  & 3967.46 & 95 $\pm$ 3 & 3588 $\pm$ 11 & 2940 $\pm$ 9 & ... & ... & ... & ... \\
H$\epsilon$ & 3970.07 & 98 $\pm$ 6 & 4016 $\pm$ 9 & 3209 $\pm$ 11 & 49 $\pm$ 4 & ... & ... & ... \\
H$\delta$ & 4101.74 & 196 $\pm$ 4 & 5757 $\pm$ 6 & 4515 $\pm$ 27 & 92 $\pm$ 4 & ... & ... & ... \\
H$\gamma$ & 4340.47 & 354 $\pm$ 5 & 9956 $\pm$ 7 & 8167 $\pm$ 18 & 142 $\pm$ 2 & ... & 62 $\pm$ 5 & ... \\
$[$OIII$]$ & 4363.21 & 91 $\pm$ 5 & 1609 $\pm$ 8 & 1399 $\pm$ 12 & 39 $\pm$ 4 & ... & ... & ... \\
H$\beta$ & 4861.33 & 779 $\pm$ 8 & 21673 $\pm$ 10 & 18180 $\pm$ 16 & 162$^a$ & 109 $\pm$ 3 & 102 $\pm$ 9 & 31 $\pm$ 2 \\
$[$OIII$]$ & 4958.91 & 1364 $\pm$ 3 & 32476 $\pm$ 16 & 30788 $\pm$ 5 & 583 $\pm$ 3 & 198 $\pm$ 2 & 158 $\pm$ 7 & ... \\
$[$OIII$]$ & 5006.84 & 4099 $\pm$ 6 & 95521 $\pm$ 16 & 92673 $\pm$ 17 & 2065 $\pm$ 3 & 492 $\pm$ 3 & 668 $\pm$ 6 & 34 $\pm$ 1 \\
$[$HeI$]$ & 5875.67 & ... & 2387 $\pm$ 13 & 2119 $\pm$ 6 & ... & ... & ... & ... \\
$[$OI$]$ & 6300.30 & ... & 851 $\pm$ 7 & 851 $\pm$ 8 & ... & ... & ... & ... \\
H$\alpha$ & 6562.82 & ... & 67942 $\pm$ 15 & 57500 $\pm$ 9 & ... & ... & ... & 102 $\pm$ 2 \\
$[$NII$]$ & 6583.41 & ... & 1356 $\pm$ 8 & 1404 $\pm$ 11 & ... & ... & ... & ... \\
$[$SII$]$ & 6716.47 & ... & 3614 $\pm$ 8 & 3302 $\pm$ 10 & ... & ... & ... & ... \\
$[$SII$]$ & 6730.85 & ... & 2638 $\pm$ 9 & 2463 $\pm$ 5 & ... & ... & ... & ... \\
$[$ArIII$]$ & 7135.78 & ... & 1302 $\pm$ 6 & 1464 $\pm$ 10 & ... & ... & ... & ... \\
\hline
\hline
\end{tabular}
\caption[]{Observed line fluxes for the primary sample. (1) Emission line;  (2) Rest-frame wavelength; (3) Observed line fluxes for ID 91; (4) Observed line fluxes for ID 891; (5) Observed line fluxes for ID 895; (6) Observed line fluxes for ID 2478; (7) Observed line fluxes for ID 2532; (8) Observed line fluxes for ID 6465; (9) Observed line fluxes for ID 7373.

$^{a}$ Uncertain value. } 
\label{tab:table_fluxes_primary}
\end{center}
\end{table*}


\section{Line fluxes of the EELG extended sample candidates}\label{sec:App_fluxes_extended}

Here, we present the fluxes and equivalent width measured with pyPlatefit for our extended sample. Only lines with SNR higher than 3 are included in the tables.

\begin{table*}
\begin{center}
\resizebox{17.cm}{!} {
\begin{tabular}{ c c | c|  c | c|  c|  c|  c  }
\hline
\hline
      &    Galaxy ID       & 1093    & 1426   & 1561  & 1699  & 6474  & 6865  \\
\hline
Line  & $\lambda$ & F & F & F & F & F & F  \\
      &    (\AA{})      &  (10$^{-20}\frac{erg}{s cm^{2}}$) & (10$^{-20}\frac{erg}{s cm^{2}}$)  &  (10$^{-20}\frac{erg}{s cm^{2}}$) & (10$^{-20}\frac{erg}{s cm^{2}}$)  &  (10$^{-20}\frac{erg}{s cm^{2}}$) & (10$^{-20}\frac{erg}{s cm^{2}}$)  \\
   (1)      &    (2)      & (3)  & (4)  & (5)  &  (6)    &  (7) & (8)   \\
\hline
$[$MgII$]$ & 2795.53 & ... & ... & ... & ... & ... & 57 $\pm$ 3 \\
$[$OII$]$ & 3726.03 & 673 $\pm$ 6 & 107 $\pm$ 4 & 164 $\pm$ 4 & 107 $\pm$ 3 & ... & 132 $\pm$ 3 \\
$[$OII$]$ & 3728.82 & 981 $\pm$ 6 & 274 $\pm$ 4 & 262 $\pm$ 4 & 102 $\pm$ 3 & ... & 126 $\pm$ 3 \\
H11 & 3770.63 & ... & ... & ... & ... & ... & 61 $\pm$ 3 \\
H10 & 3797.90 & ... & ... & ... & ... & ... & 78 $\pm$ 3 \\
H9 & 3835.39 & ... & ... & ... & ... & ... & 102 $\pm$ 2 \\
$[$NeII$]$ & 3868.75 & ... & 175 $\pm$ 3 & 135 $\pm$ 4 & 52 $\pm$ 3 & ... & 382 $\pm$ 2 \\
$[$HeI$]$ & 3888.65 & ... & 98 $\pm$ 3 & ... & ... & ... & 214 $\pm$ 2 \\
H8 & 3889.05 & ... & 112 $\pm$ 4 & 63 $\pm$ 4 & ... & ... & 220 $\pm$ 2 \\
$[$NeII$]$ & 3967.46 & 185 $\pm$ 6 & ... & ... &  & ... & 166 $\pm$ 3 \\
H$\epsilon$ & 3970.07 & 243 $\pm$ 7 & 93 $\pm$ 4 & ... &  & ... & 222 $\pm$ 4 \\
H$\delta$ & 4101.74 & 489 $\pm$ 8 & 154 $\pm$ 4 & 103 $\pm$ 5 & 55 $\pm$ 4 & ... & 347 $\pm$ 3 \\
H$\gamma$ & 4340.47 & 620 $\pm$ 7 & 276 $\pm$ 4 & 175 $\pm$ 5 & ... & 2260 $\pm$ 7 & 659 $\pm$ 3 \\
$[$OIII$]$ & 4363.21 & 75 $\pm$ 6 & ... & ... & ... & 605 $\pm$ 6 & 208 $\pm$ 2 \\
H$\beta$ & 4861.33 & 1347 $\pm$ 9 & 626 $\pm$ 4 & 352 $\pm$ 6 & 221 $\pm$ 5 & 5023 $\pm$ 6 & 1453 $\pm$ 3 \\
$[$OIII$]$ & 4958.91 & 1807 $\pm$ 7 & 771 $\pm$ 3 & 484 $\pm$ 4 & 309 $\pm$ 4 & 8562 $\pm$ 7 & 2054 $\pm$ 2 \\
$[$OIII$]$ & 5006.84 & 5364 $\pm$ 9 & 2485 $\pm$ 5 & 1450 $\pm$ 6 & 905 $\pm$ 5 & 26044 $\pm$ 8 & 6276 $\pm$ 3 \\
$[$HeI$]$ &  5875.67 & 135 $\pm$ 7 & 53 $\pm$ 3 & ... & ... & 543 $\pm$ 7 & ... \\
$[$OI$]$ &  6300.30 & ... & ... & .... & ... & 156 $\pm$ 6 & ... \\
H$\alpha$ &  6562.82 & ... & ... & .... & ... & 15343 $\pm$ 7 & ... \\
$[$NII$]$ &  6583.41 & ... & ... & .... & ... & 214 $\pm$ 6 & ... \\
$[$SII$]$ &  6716.47 & ... & ... & ... & ... & 588 $\pm$ 6 & ... \\
$[$SII$]$ &  6730.85 & ... & ... & ... & ... & 444 $\pm$ 6 & ... \\
$[$ArIII$]$ & 7135.78 & ... & ... & ... & ... & 293 $\pm$ 7 & ... \\
\hline
\hline
\end{tabular}
}
\caption[]{Same as Table \ref{tab:table_fluxes_primary} for the extended sample. $^{a}$ Uncertain value.}
\label{tab:table_fluxes_extended}
\end{center}
\end{table*}


\end{appendix}

\end{document}